\documentclass[12pt]{article}

\usepackage{graphicx}

\def\pw#1{^{#1}}

\def\beeq{\begin{eqnarray}} \def\eeeq{\end{eqnarray}}
\newcommand\mysection{\setcounter{equation}{0}\section}
\renewcommand{\theequation}{\thesection.\arabic{equation}}
\newcounter{hran} \renewcommand{\thehran}{\thesection.\arabic{hran}}

\def\bmini{\setcounter{hran}{\value{equation}}\refstepcounter{hran}\setcounter{equation}{0}\renewcommand{\theequation}{\thehran\alph{equation}}\begin{eqnarray}}
\def\bminiG#1{\setcounter{hran}{\value{equation}}\refstepcounter{hran}\setcounter{equation}{-1}\renewcommand{\theequation}{\thehran\alph{equation}}\refstepcounter{equation}\label{#1}\begin{eqnarray}}

\def\emini{\end{eqnarray}\relax\setcounter{equation}{\value{hran}}\renewcommand{\theequation}{\thesection.\arabic{equation}}}

\def\fun#1#2{\lower3.6pt\vbox{\baselineskip0pt\lineskip.9pt
  \ialign{$\mathsurround=0pt#1\hfil##\hfil$\crcr#2\crcr\sim\crcr}}}

\def\eV{{\rm e\kern-0.12em V}}            
 \def\GeV{{\rm G}\eV} 
\def\half{{\textstyle {1\over2}}}
  
\def\fm{\,{\rm fm}}
\def \al {\relax\ifmmode{\alpha}\else{$\alpha${ }}\fi}

\def\ben{\begin{enumerate}}  \def\een{\end{enumerate}}
\def\bit{\begin{itemize}}    \def\eit{\end{itemize}}
\def\beq{\begin{equation}}   \def\eeq{\end{equation}}
\def\bea{\begin{eqnarray}}  \def\eea{\end{eqnarray}}
\def\nn{\nonumber}
\def\noi{\noindent}

\def\kp{\relax\ifmmode{k_\perp}\else{$k_\perp${ }}\fi}
\def\kps{\relax\ifmmode{k_\perp\pw2}\else{$k_\perp\pw2${ }}\fi}
\def \as{\relax\ifmmode\alpha_s\else{$\alpha_s${ }}\fi}

\newskip\humongous \humongous=0pt plus 1000pt minus 1000pt
\def\caja{\mathsurround=0pt}

\newif\ifdtup

\def\eqal2#1{\,\vcenter{\openup1\jot
\caja   \ialign{\strut \hfil$\displaystyle{##}$&\hfil$
\displaystyle{{}##}$\hfil &$
\displaystyle{{}##}$\hfil\crcr#1\crcr}}\,}

\def\lsim{\raise0.3ex\hbox{$<$\kern-0.75em\raise-1.1ex\hbox{$\sim$}}}
\def\gsim{\raise0.3ex\hbox{$>$\kern-0.75em\raise-1.1ex\hbox{$\sim$}}}
 \def\cite#1{[\ref{#1}]}
 \def\citd#1#2{[\ref{#1},\ref{#2}]}
 
 \def\citm#1#2{[\ref{#1}--\ref{#2}]}

\def\beql#1{\beq\label{#1}}

\def\np#1#2#3{{\em Nucl.~Phys.}~\underline{B#1} (19#3) #2}
\def\pl#1#2#3{{\em Phys.~Lett.}~\underline{#1B} (19#3) #2}
\def\prD#1#2#3{{\em Phys.~Rev.}~\underline{D#1} (19#3) #2}
\def\pr#1#2#3{{\em Phys.~Rev.}~\underline{#1} (19#3) #2}

\def\prl#1#2#3{{\em Phys.Rev.Lett.}~\underline{#1} (19#3) #2}

\begin{document}

\thispagestyle{plain}
\setcounter{page}{1}
\vbox to 1 truecm {}

\begin{flushright}
May 1996 \\[0.1cm]
BI-TP 96/21\\
CUTP--759\\
LPTHE-Orsay 96-34
 \\
\end{flushright}

\vfill
\def\cen{\centerline}

\cen{{\bf \large RADIATIVE ENERGY LOSS OF HIGH ENERGY QUARKS AND}}
\cen{{\bf \large GLUONS IN A FINITE VOLUME QUARK-GLUON PLASMA}}
\renewcommand{\thefootnote}{\fnsymbol{footnote}}
\vskip 1.5 truecm
\centerline
{\bf R.~Baier~$\pw1$, Yu.~L.~Dokshitzer~$\pw2$, 
A.~H.~Mueller\footnote[1]{Supported in  part
by the U.S. Department of Energy under grant DE-FG02-94ER-40819}~$\pw{3}$$^{,}$ $\pw4$,
S.~Peign\'e~$\pw3$ and D.~Schiff~$\pw3$} \vskip 10 pt 
\centerline{{\it $\pw{1 }$Fakult\"at
f\"ur Physik, Universit\"at Bielefeld, D-33501 Bielefeld, Germany}}
\centerline{{\it $\pw{2 }$ Theory Division, CERN, 1211 Geneva 23,
Switzerland\footnote[2]
{Permanent address: Petersburg Nuclear Physics Institute, Gatchina,
 188350 St. Petersburg, Russia } }}
\centerline{{\it $\pw{3 }$
LPTHE\footnote[3] { Laboratoire associ\'e au
Centre National de la Recherche Scientifique - URA D0063}
, Universit\'e  Paris-Sud, B\^atiment 211, F-91405 Orsay, France}}
\centerline{{\it $\pw{4 }$Physics Department, Columbia University,
New York, NY 10027, USA\footnote[4] {Permanent address}}}

\renewcommand{\thefootnote}{\arabic{footnote}}
\vskip 2 cm
\cen{\bf Abstract}
\vskip 5pt
\noindent
{The medium induced energy loss spectrum of a high energy quark or gluon traversing 
a hot QCD medium of finite volume is studied. 
We model the interaction by a simple picture of static scattering centres. 
The total induced energy loss is found to grow as $L^2$, where $L$
is the extent of the medium. The solution of the energy loss problem is reduced 
to the solution of a Schr\"odinger-like equation whose ``potential'' 
is given by the single-scattering cross section of the high energy parton 
in the medium. 
These results should be directly applicable to a quark-gluon plasma.}

\vfill \eject

\mysection{Introduction}
\label{sec:nom1}
The determination of the radiative energy loss of a high energy charged particle 
as it passes through matter is a problem studied some time ago, in QED, by Landau, 
Pomeranchuk and Migdal \citm{LP}{Ter}. 
There is recent data from SLAC \cite{SLAC} on radiative energy loss in QED \cite{BD}.
New interest in this problem \citm{GW}{BDMPS} has arisen because
the corresponding problem in QCD, that of the energy loss of a high energy quark 
or gluon due to medium stimulated gluon radiation, may be important as a signal 
for quark-gluon plasma formation in high energy heavy ion collisions. 

Recently this problem was considered for infinite matter in \cite{BDPS}, hereinafter
referred to as BDPS. Using the Gyulassy-Wang (GW) model \cite{GW} for hot matter they
observed that the QED and QCD problems are mathematically equivalent 
if one identifies the emission angle of radiated photons in the QED case 
with the transverse momentum of radiated gluons in the QCD case.
The energy loss per unit length, $- \displaystyle{{dE/dz}}$, 
in hot QCD matter was found to be proportional to $\alpha_s \sqrt{E}$ for an
incident parton of energy $E$. The $\sqrt{E}$ growth of 
$-\displaystyle{{dE/dz}}$ was unexpected and suggested that energy losses 
of high energy jets in a quark-gluon plasma might be large. 
The 
procedure used by BDPS was not adequate for
determining the exact logarithmic prefactor in $-\displaystyle{{dE/dz}}$. 
The problem was revisited in \cite{BDMPS}, hereinafter referred to as BDMPS, 
where a simple differential equation was given to determine the spectrum 
$\omega \displaystyle{{dI/d \omega dz}}$ for radiated photons (in QED) and gluons
(in QCD) which corrects the prefactor of the result in BDPS. \par

The present paper generalizes the BDMPS approach to finite length hot matter. 
Although our discussion is carried out in the context of QCD it is a simple matter 
to change variables in the QCD results to get the corresponding QED results. 
Our main interest is in the situation where the incident quark or gluon 
is sufficiently energetic so that the length of the matter, $L$, satisfies 
$L < L_{cr} = \sqrt{\lambda _g E / \mu^2}$, with $\lambda_g$ 
the gluon mean free path and $\mu$ the Debye screening mass of the medium. 
Our principal results are for 
$\omega \displaystyle{{dI/d\omega dz}}$, given in (\ref{DiffSpectrum}), 
and for $dE/dz$ given in (\ref{Loss1}) and (\ref{Loss2}). 
The total energy loss $\Delta E$ in hot QCD matter of length $L$ is 
$\Delta E = \displaystyle{{\alpha_s C_R \over 8}} \displaystyle{{\mu^2 \over
\lambda_g}} L^2 \ln (L/\lambda_g)$ 
as given in (\ref{TotalLoss}) with $R$ the colour representation of the incident
parton. The total energy loss is found to be proportional to $L^2$, 
a result which is surprising at first glance. 
However, by considering the limiting case where $L = L_{cr}$ it becomes clear 
that this corresponds to the BDPS result $\Delta E \propto L_{cr} \sqrt{E}$. 
The fact that $\Delta E$ is proportional to $L^2$ 
has a simple physical interpretation. 
Since we expect the energy weighted spectrum 
$\omega \displaystyle{{dI/d \omega dz}}$ to be integrable in the infrared, 
$\Delta E$ is roughly determined by the maximum energy a radiated gluon can
have still maintaining a coherence length $\leq L$. 
But, the formation time of a radiated gluon is 
$\displaystyle{{2 \omega/ k_{\bot max}^2}}$, with $k_{\bot max}$ 
the maximum transverse momentum that the gluon gets by rescattering in the medium 
as it is being produced. 
Taking $k_{\bot max}^2 = \displaystyle{{L\mu^2 / \lambda_g}} $, with $\mu$ the
typical momentum transfer to the gluon in a single scattering, 
and setting 
$L \propto \displaystyle{{2 \omega \over (L/\lambda_g)\mu^2}}$  one finds 
$\omega \propto \displaystyle{{1 \over 2}}\displaystyle{{\mu^2 \over \lambda_g}} L^2$.
This estimate, given more precisely in (\ref{TypicalEnergy}), leads to the 
$L^2$ dependence of $\Delta E$.  

The outline of the paper is as follows:

In section \ref{sec:nom2} the emission probability of a soft gluon from a high energy
quark traversing hot QCD matter is given in the GW model. The basic emission vertex is
calculated in detail as are the subsequent rescatterings of the quark-gluon system
passing through the matter as the gluon is becoming free. While, for simplicity, much
of the discussion is given in the large-$N_c$ limit, final formulas are given with
exact colour factors. After observing that the QED and QCD emission formulas are
identical, with the identification of corresponding variables, the formula for the
radiation intensity spectrum $\omega \displaystyle{{dI / d \omega dz}}$ for
infinite volume hot QCD matter is given as a direct consequence of the QED spectrum
derived by BDMPS. \par

In section \ref{sec:nom3} a heuristic discussion of energy loss is given both for 
$L \ll L_{cr}$ and for $L \gg L_{cr}$. By requiring that these results match 
at $L_{cr}$ a direct connection between $\Delta E \propto L \sqrt{E}$ 
for $L \gg L_{cr}$ and $\Delta E \propto L^2$ for $L \ll L_{cr}$ is made. \par

Section \ref{sec:nom4} is concerned with deriving the general equations governing
radiative energy loss in a hot QCD plasma of extent $L$. 
The basic equation determining energy loss 
is a Schr\"odinger-like equation whose ``potential'' is
given in terms of a single scattering cross section, in impact parameter space, of a
high energy parton. 
We expect this same formalism, but with a different ``potential'', 
to apply to the energy loss problem in cold nuclear
matter \cite{BDMPS3}. We presume, however, that the magnitude of the energy loss in hot
and cold matter may be quite different. \par

In section \ref{sec:nom5} we give an approximate solution for the radiation spectrum
valid, at large $L$, for those gluon energies dominating the energy loss of the
primary parton. Our main result is given in (5.16). We note that as 
$L \to \infty$
the spectrum agrees with the result previously given in BDMPS. \par

Formulas for the total energy loss due to medium induced radiation are given in section
\ref{sec:nom6}. We expect the most likely place that these results may have direct
phenomenological application is in high energy jet production when a quark-gluon
plasma is formed in heavy ion collisions. We also expect results much
like (6.6) to hold in jet production in cold nuclear matter, but that is the subject
for a subsequent paper \cite{BDMPS3}. \par

Gluon emission with double scattering is given in Appendix A while rules for dealing
with colour factors for multiple scattering are given in Appendix B.
The calculation of the planar diagrams necessary to obtain (2.31) is given
in Appendix C. A curious integral which arises in evaluating the energy loss is
calculated in Appendix D.       

 \mysection{General expression of the medium
induced radia\-tion spectrum in QCD} \label{sec:nom2} 
Here we derive the general form of the
gluon energy spectrum induced by the propagation of a high energy quark 
in a finite non-abelian medium. Results are given for the case of an incident parton
of arbitrary colour representation $R$.
We also establish a formal analogy between QED and QCD.

\subsection{Model for multiple scattering}

In order to describe the successive interactions of a high energy incident parton with a
hot QCD medium, we use the model introduced by Gyulassy and Wang  
and recently used in BDMPS 
to study the photon energy spectrum induced by multiple QED scattering of a fast charged
particle. The main feature of the model consists in assuming that
scattering centres are static. This allows one to focus on purely radiative processes, since
the collisional energy loss then vanishes. The centre located at $\vec{x}_i$ creates a
screened Coulomb potential 
\beq {\cal V}_i(\vec{x}) = {g \over 4 \pi} \ {e^{-\mu |\vec{x} -
\vec{x}_i|} \over |\vec{x} - \vec{x}_i|} \ \ \ , \label{potentialx} 
\eeq
with Fourier transform
\beq
{\cal V}_i(\vec{q}) = {g \over \vec{q}^{\ 2} + \mu^2} \ e^{-i\vec{q} \cdot \vec{x}_i} \ \ \ ,
\label{potentialq} 
\eeq
where $g$ is the QCD coupling constant. 
We suppose that the range of
the potentials ${\cal V}_i$ is small compared to the mean free path $\lambda$ of the
incident parton,
\beq
\mu^{-1} \ll \lambda \ \ \ , \label{mulambda}
\eeq
\noi where $\mu$ is the Debye mass induced by the medium. This means that successive
scatterings are independent, since the incident parton cannot scatter simultaneously
off two distinct centres. As a consequence, its propagation is ``time-ordered'', and
we may number the scattering centres according to the interaction time (or equivalently
the longitudinal coordinate) of the radiating parton. (See \cite{BDMPS} for a
justification in terms of Feynman diagrams). Moreover, in the context of one gluon
emission, this assumption allows us to neglect amplitudes involving
four-gluon vertices of the type shown in Fig.~1. 

\begin{figure}[h]
\centering
\includegraphics[height=3cm]{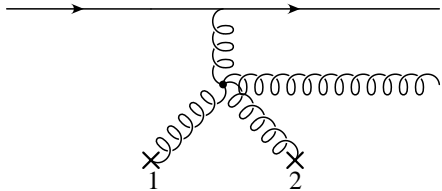}
\caption{\it{This gluon emission amplitude induced by the scattering of an energetic quark off two static centres 1 and 2 is negligible in the $\lambda \gg \mu^{-1}$ limit.}}
\label{fig1}
\end{figure}

\noi Finally, we work in the limit of very high energy $E$ for the incident parton and in
the soft gluon approximation, \beq
\omega \ll E \ \ \ . \label{soft}
\eeq
\noi Let us exhibit the implications of these two assumptions by giving the basic
emission amplitude in a single scattering.

\vskip 5 truemm
\subsection{Gluon emission induced by a single scattering}
\vskip 5 truemm
The emission amplitude is depicted in Fig.~2. It includes the emission off the
projectile (from now on chosen to be a quark) given by $M_1$ and the emission off the
exchanged gluon given by $M_2$. \par

\begin{figure}[h]
\centering
$M_1=
\begin{minipage}[b]{.25\textwidth}
\centering
\includegraphics[width=3.5cm]{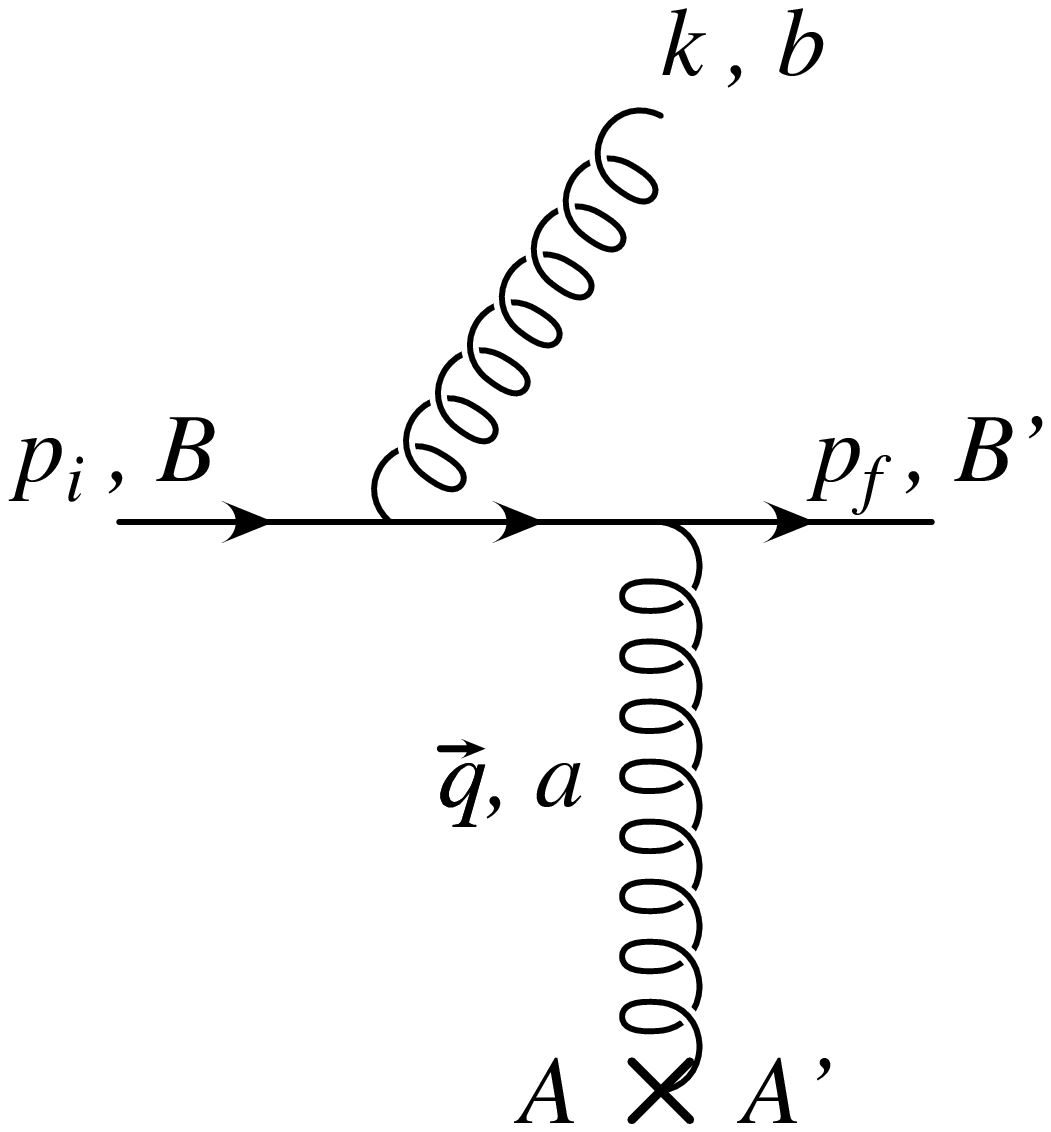}
\par\vspace{-50pt}
\end{minipage}
+
\begin{minipage}[b]{.25\textwidth}
\centering
\includegraphics[width=3.5cm]{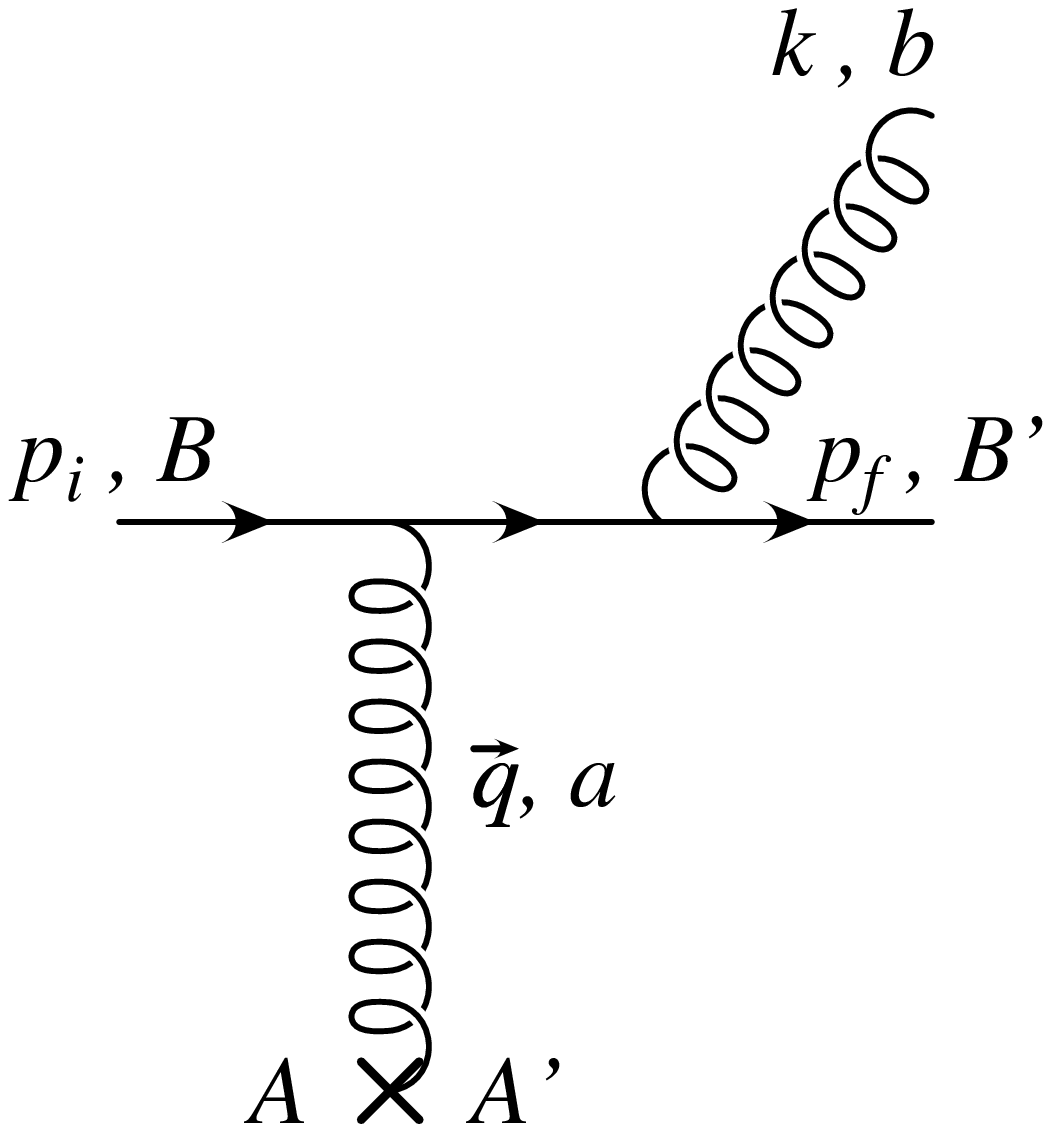}
\par\vspace{-50pt}
\end{minipage}
\ ;\ M_2=
\begin{minipage}[b]{.25\textwidth}
\centering
\includegraphics[width=3.5cm]{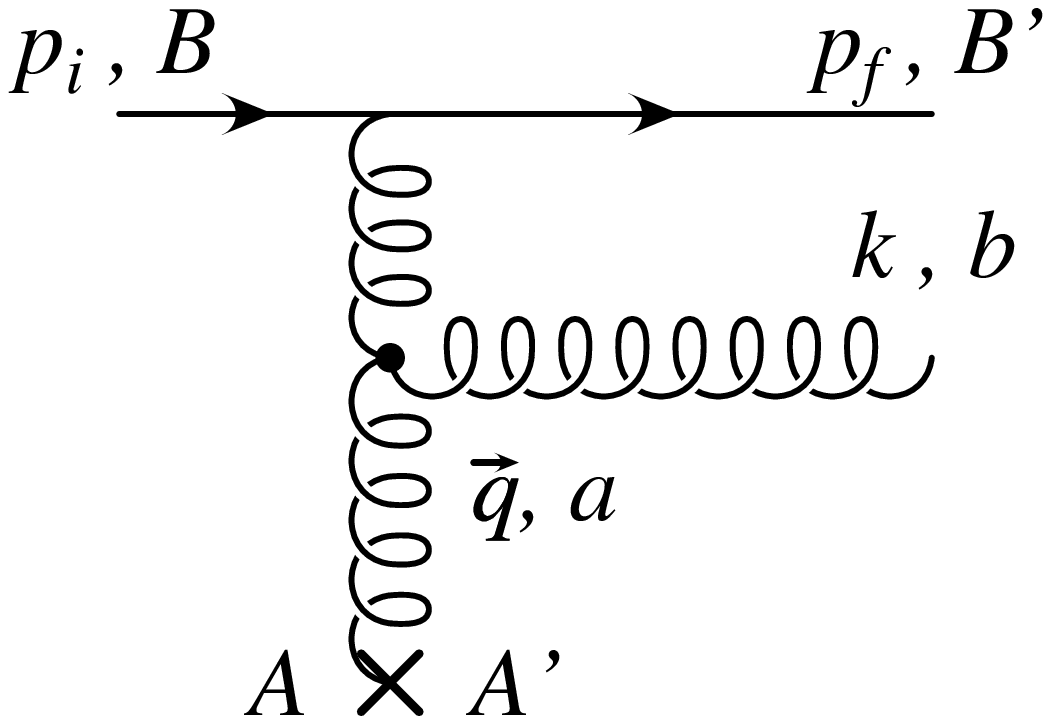}
\par\vspace{-50pt}
\end{minipage}$
 
\vspace{50pt}
\caption{\it{Gluon emission amplitude induced by one scattering.}}
\label{fig2}
\end{figure}

\noi The colour indices of the static centre and of the incident quark are denoted by
$A$, $A'$ and $B$, $B'$ respectively. The indices of the exchanged and radiated gluons
are $a$ and $b$. Neglecting screening for the moment, we write the amplitude for elastic
scattering off a static source as
\bminiG{EDh}
 M_{e\ell} = T_{B'B}^a \ M_{A'A}^a \ \ ; \ \ M_{A'A}^a = g_{\mu \nu} \ M_{A'A}^{a, \mu
\nu} \> . \label{elastic1}
\eeeq
Here
\beeq
M_{A'A}^{a, \mu \nu} = ig^2 (p_i + p_f)^{\mu} {1 \over q^2} \ \delta_0^{\nu}
\ T_{A'A}^a  \> , \label{elastic2}
\emini
where we neglected spin effects in the high energy limit.
The static source can be viewed as if it were a heavy quark.

In Feynman gauge, the amplitude $M_1$ (Fig.~2) for soft gluon emission may be
expressed as the elastic scattering amplitude times a radiation factor
as 
\beq
M_1 \simeq - g \left \{ {\varepsilon \cdot p_f \over k \cdot p_f} (T^b T^a)_{B'B}
-  {\varepsilon \cdot p_i \over k \cdot p_i} (T^a T^b)_{B'B} \right \} M_{A'A}^a \
\ \ , \label{M1}  \eeq
\noi where $\varepsilon$ denotes the gluon polarization state.
The generators of the fundamental representation of $SU(N_c)$ are $T^a(a = 1, 
\ldots N_c^2 -1)$, satisfying $[T^a , T^b] = i f^{abc}\ T^c$. 
In the same way we get
\beq
M_2 \simeq - g {2 \over (p_f - p_i)^2} \left \{ g_{\mu \nu}  \varepsilon \cdot (p_f
-  p_i) - k_{\mu} \varepsilon_{\nu} + k_{\nu} \varepsilon_{\mu} \right \} \cdot
M_{A'A}^{a, \mu \nu} \ [T^b , T^a]_{B'B} \ . \label{M2} 
\eeq
In addition to $M_1$ and $M_2$, there is a term $M_3$ coming from
gluon radiation off the static source. The sum of the three terms 
is gauge invariant. In a physical gauge such
as light-cone gauge, $M_3$ is down by a factor of $k_\perp/\omega$ compared to $M_1$
and $M_2$. In the calculation given below we use light-cone gauge and assume
 $k_\perp/\omega\ll 1$.

In a hot plasma the source is screened
as indicated by (\ref{potentialx}) and (\ref{potentialq}) in the GW model. The
reader may have doubts as to the general gauge invariance of that model. These doubts may be
put to rest by the following arguments. It is straightforward to show that $M_1 + M_2 + M_3$
remains gauge invariant when the emitted and exchanged gluons are given the same mass $\mu$.
As we shall see later, the emitted gluon has a small impact parameter for the physical
problem we consider. As a consequence of the small impact parameter, one may neglect
the mass for the emitted gluon; keeping the mass $\mu$ only for the exchanged gluon
leads to the Gyulassy-Wang model. 

In light-cone gauge 
\beq
\varepsilon = \left ( \varepsilon_0, - \varepsilon_0 , \vec{\varepsilon}_{\bot}
\right ) ; \ \varepsilon \cdot k = 0 \Rightarrow \varepsilon_0 =
{\vec{\varepsilon}_{\bot} \cdot \vec{k}_{\bot} \over \omega + k_{//}}  \simeq
{\vec{\varepsilon}_{\bot} \cdot \vec{k}_{\bot} \over 2 \omega}  \ . \label{polarization} 
\eeq 
In the high energy limit
\beq
{\varepsilon \cdot p_i \over k \cdot p_i} \simeq {\varepsilon \cdot p_f \over k \cdot p_f}
\simeq 2 \vec{\varepsilon}_{\bot} \cdot {\vec{k}_{\bot} \over k_{\bot}^2} \>,
\label{kinematics} 
\eeq 
where $\vec{k}_{\bot}$ is the transverse momentum of the gluon with respect to the
direction of the incident particle. Thus,
\beq
M_1 \simeq - 2g \ \vec{\varepsilon}_{\bot} \cdot {\vec{k}_{\bot} \over k_{\bot}^2} \ [T^b ,
T^a]_{B'B} \ M_{A'A}^a \ . \label{M1bis}
 \eeq
In QED \cite{BDMPS}, the photon radiation amplitude vanishes in the limit $E \to
\infty$. In QCD, in the high energy limit only the purely non-abelian contribution to
the gluon radiation spectrum survives. This is underlined by the presence of the
commutator in (\ref{M1bis}).
As a result we can use the eikonal approximation where the
trajectory of the projectile is taken to be a straight line. Also, 
\beq
M_2 \simeq 2 g \ \vec{\varepsilon}_{\bot} \cdot {\vec{k}_{\bot} - \vec{q}_{\bot} \over (
\vec{k}_{\bot} - \vec{q}_{\bot})^2} \ [T^b , T^a]_{B'B} \ M_{A'A}^a \ . \label{M2bis} 
\eeq
Finally, the radiation amplitude induced by one scattering of momentum transfer
$\vec{q}$ reads 
\beq
M_1 + M_2 \simeq - 2g \ \vec{\varepsilon}_{\bot} \cdot \vec{J}(k, q) \ [T^b , T^a]_{B'B}
\ M_{A'A}^a \ \ \ , \label{M1+M2} \eeq
\noi where the emission current $\vec{J}$ is defined as 
\beq
\vec{J}(k, q) = {\vec{k}_{\bot} \over k_{\bot}^2} - {\vec{k}_{\bot} - \vec{q}_{\bot}
\over ( \vec{k}_{\bot} - \vec{q}_{\bot})^2} \ \ \ . \label{current} 
\eeq
We are interested in the gluon energy spectrum, which is given by the ratio between
the radiation and elastic cross sections. 
Up to a common flux factor
\bminiG{q-def1}
&&d\sigma_{e\ell} \propto C_F |M^a|^2 \ dq_{\bot}^2 \ ; \ C_F = {N_c^2 - 1 \over 2 N_c} \ \
\ ,  \label{elcrossection} \\
&&d\sigma_{rad} \propto {\alpha_s \over \pi^2} \ C_F N_c  \ \vec{J}^{\,2} \ |M^a|^2 \
d^2\vec{k}_{\bot} {dk_{//} \over \omega} dq_{\bot}^2 \ \ \ , 
\label{radcrosssection} 
\emini
where $|M^a|^2$ is defined as 
$|M ^a|^2 \delta^{a'a} = \displaystyle{{1 \over N_c}} {\rm Tr}\, ( M^a M^{a'})$. 
Thus we obtain, for $|\vec{k}_{\bot}| \ll \omega \ll E$,
\beq
\omega {dI \over d \omega d^2 \vec{k}_{\bot}} = N_c \ {\alpha_s \over \pi^2}
\left < \vec{J}(k, q)^2 \right > \ \ \ . \label{spectrum1} \eeq
\noi As the amplitude has been evaluated for a fixed momentum transfer 
$\vec{q}_{\bot}$, an average over $\vec{q}_{\bot}$ has to be performed. For this we use
the probability density deduced from the elastic scattering cross section which is
easily obtained from (\ref{potentialq}). 
Thus in (\ref{spectrum1}) we define
\beq
\left < \ (\ \ ...\ \ )\ \right >\> \equiv\> 
\int d^2 \vec{q}_{\bot} \ V(q_{\bot}^2)
(\ \ ...\ \ ) \> , \label{averageq} 
\eeq
where the normalized cross section for elastic quark scattering reads
\beq
V(q_{\bot}^2) = {\mu^2 \over \pi (q_{\bot}^2 + \mu^2)^2} \>, \label{Potential}
\eeq
with $\int d^2\vec{q}_{\bot} \ V(q_{\bot}^2) = 1$. We have used the fact that the
longitudinal transfer is negligible with respect to $q_{\bot}$ when $E \to \infty$. As we
aim to derive the radiation density induced by multiple scattering, it is convenient to keep
the colour structure together with the current and introduce
\beq
\vec{J}_{eff} = \vec{J} (k, q) \ [T^b , T^a]_{B'B} \ \ \ . \label{EffCurrent1}
\eeq
\noi The fact that the colour structure is the same as the three-gluon vertex allows
one to give a compact diagrammatic representation of the effective current as shown in Fig.~3. 

\begin{figure}[h]
\label{fig3}
\centering
$
\begin{minipage}[b]{.25\textwidth}
\centering
\includegraphics[width=4cm]{fig3a.ps}
\par\vspace{-50pt}
\end{minipage}
+
\begin{minipage}[b]{.25\textwidth}
\centering
\includegraphics[width=4cm]{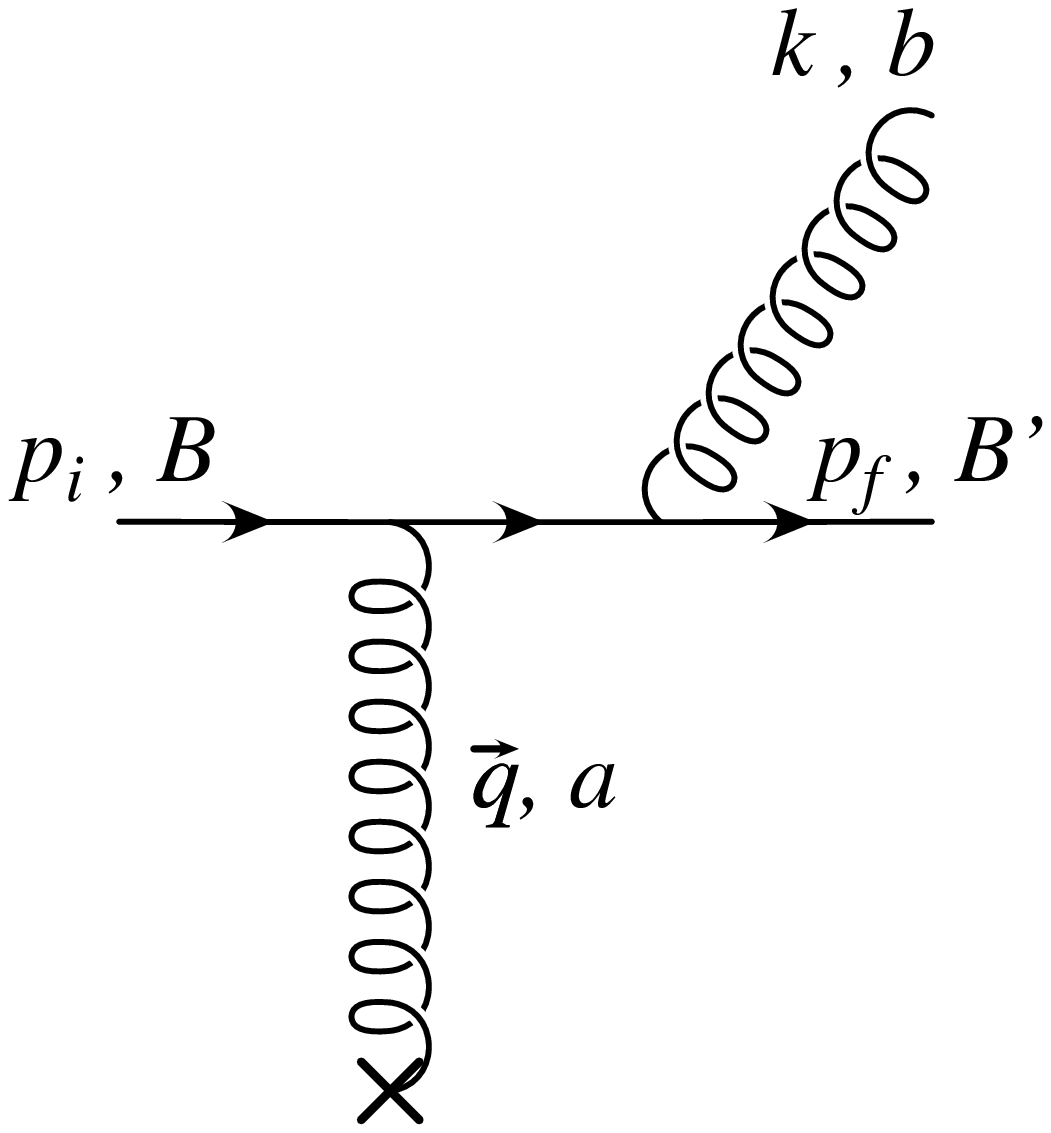}
\par\vspace{-50pt}
\end{minipage}
+
\begin{minipage}[b]{.25\textwidth}
\centering
\includegraphics[width=4cm]{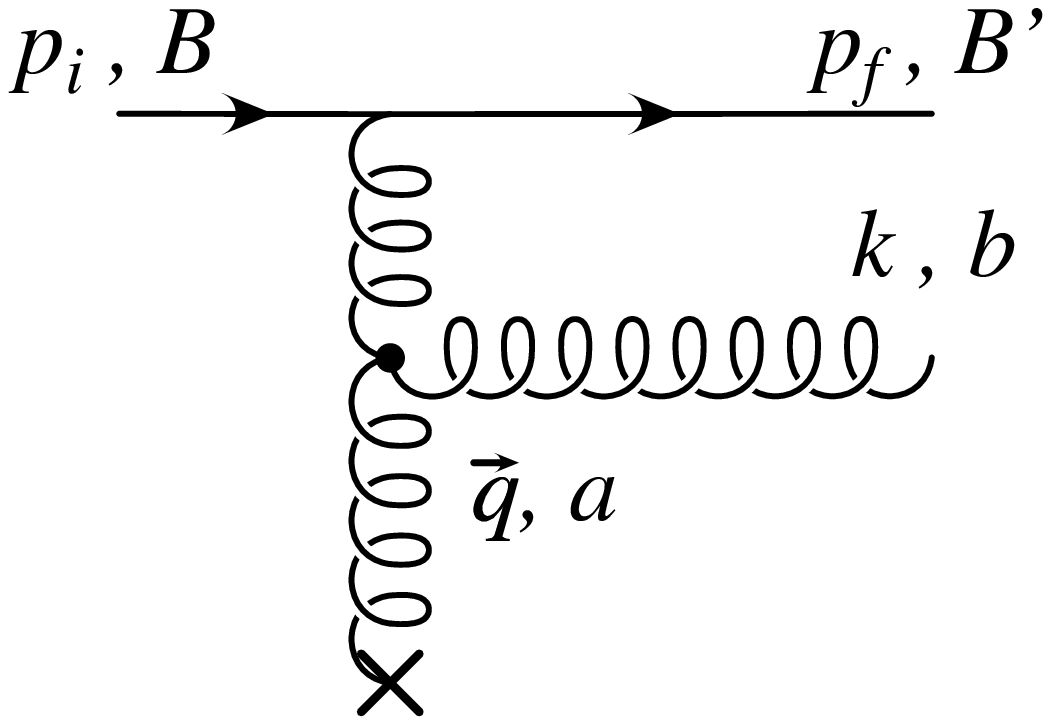}
\par\vspace{-50pt}
\end{minipage}
$ \vskip 2cm
$\equiv
\begin{minipage}[c]{.5\textwidth}
\centering 
\includegraphics[width=5cm]{fig3d.ps}
\end{minipage}
$
\caption{\it{Diagrammatic representation of the effective current (\ref{EffCurrent1}).}}
\end{figure}

Then the differential energy spectrum is simply written as
\beq
\omega {dI \over d\omega \ d^2\vec{k}_{\bot}} = {\alpha_s \over \pi^2} \ 
{\left < |\vec{J}_{eff}|^2 \right > \over C_F \ N_c} \ \ \ . \label{spectrum2} 
\eeq
A comparison between (\ref{M1+M2}) and (\ref{EffCurrent1}) allows one 
to set the proper colour factor in order to normalize to the 
elastic scattering cross section. 
The square $|\vec{J}_{eff}|^2$ includes the sum over all colour indices and the 
$N_c$ in the denominator cancels the sum over initial quark colours while 
$C_F$ corresponds to the colour factor of the normalizing elastic cross section. 
We see that the spectrum (\ref{spectrum2}) has exactly
the same form as in QED \cite{BDMPS}, up to the replacement of the photon angle by the
gluon transverse momentum (and up to colour factors). \par

The introduction of the effective current given in (\ref{EffCurrent1}) or in Fig.~3 will
provide an important simplification in the case of multiple scattering. Let us indicate how
this simplification appears in the case of two scatterings.
\vskip 5 truemm

\subsection{Effective radiation amplitudes for double and multiple scat\-te\-ring} 
\paragraph{Double scattering.}

For two scatterings, the radiation amplitude is given by a collection 
of seven diagrams.
These are simply calculated in the framework of time-ordered perturbation theory. We
show in Appendix A that all amplitudes may be grouped in effective radiation amplitudes
induced by momentum transfers $\vec{q}_1$ or $\vec{q}_2$ at times $t_1$ and $t_2$~; 
each is associated with a corresponding phase factor 
\bea
M_{rad} \propto && \Big \{ \vec{J} (k, q_1) b[c, a] \ e^{it_1 {k_{\bot}^2 \over 2
\omega}} \nn \\
&&+ \vec{J} (k, q_2) [c, b] a \ e^{it_2 {k_{\bot}^2 \over 2 \omega}} \nn \\
&&+ \vec{J} (k - q_2, q_1) [[c, b], a] \ e^{it_1 {(k - q_2)_{\bot}^2 \over 2 \omega} +
it_2 {k_{\bot}^2 - (k - q_2)_{\bot}^2 \over 2 \omega}} \Big \} \label{Mrad1} \eea
\noi (see Appendix A for the notation
concerning colour factors). This expression multiplies the elastic double scattering
amplitude and may be represented diagrammatically as an effective emission 
current as in Fig.~3. This is shown in Fig.~4.

\begin{figure}[h]
\label{fig4}
\centering
$M_{rad}\propto\left (
\begin{minipage}[c]{.2\textwidth}
\centering
\includegraphics[width=3cm]{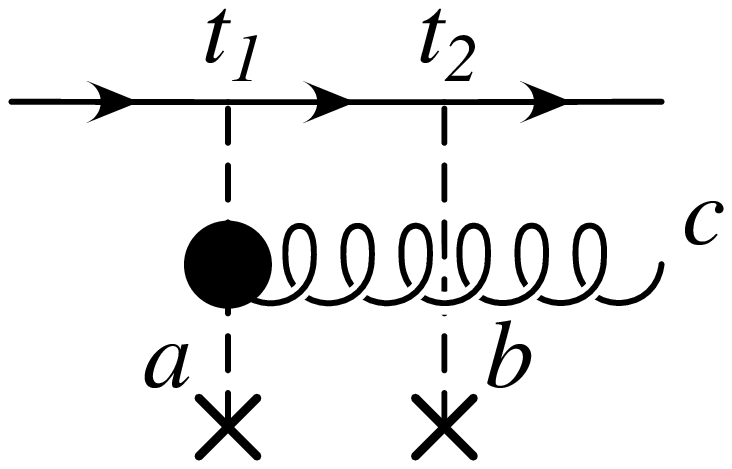}
\end{minipage}
\right )\ e^{i\phi_1}\ +\left (
\begin{minipage}[c]{.2\textwidth}
\centering
\includegraphics[width=3cm]{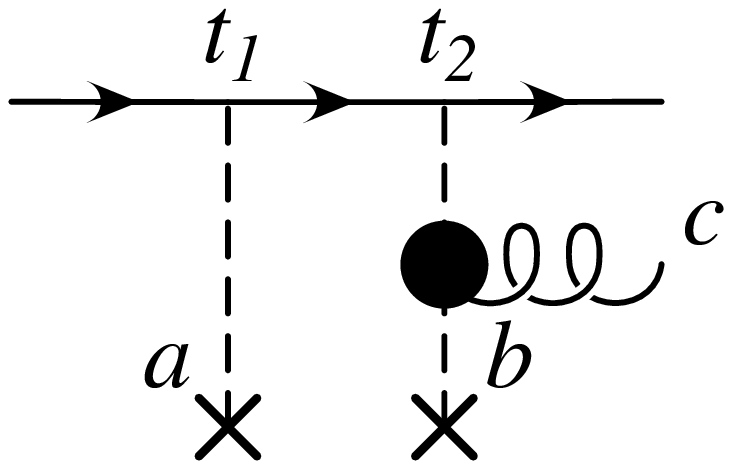}
\end{minipage}
\right )\ e^{i\phi_2}\ +\left (
\begin{minipage}[c]{.2\textwidth}
\centering
\includegraphics[width=3cm]{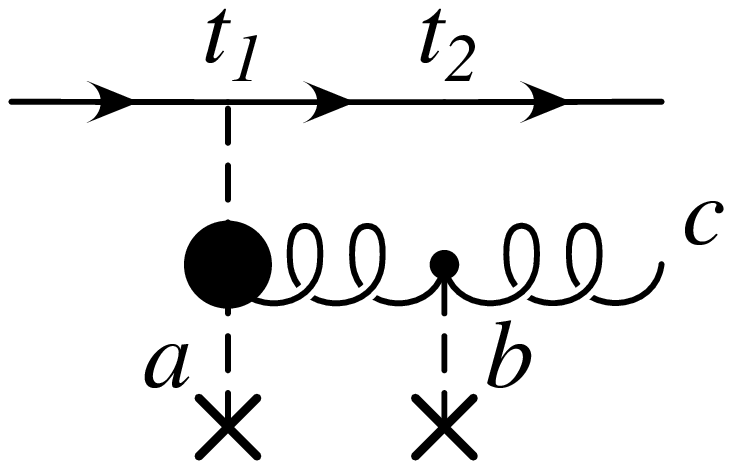}
\end{minipage}
\right )\ e^{i\phi_{12}}
$

\caption{\it{Radiation amplitude expressed in terms of effective currents.}}
\end{figure}

We thus have
\beq
M_{rad} \propto \left \{ \vec{J}_{eff}(k, q_1) e^{i \varphi_{_1}} + \vec{J}_{eff}(k, q_2)
e^{i \varphi_{_2}} + \vec{J}_{eff} (k - q_2, q_1) \ e^{i \varphi_{_{12}}} \right \} \ \ \ ,
\label{Mrad2} \eeq
\noi where we use an obvious notation for the phases. 
The first term on the right-hand side
of (\ref{Mrad2}) and Fig.~4 corresponds to gluon emission at $t_1$ followed 
by rescattering of the quark at $t_2$. The second is gluon production at $t_2$ while the third is gluon
production at $t_1$ followed by rescattering of the gluon at $t_2$. 
As seen from (\ref{Mrad2}),
quark rescattering does not affect the phase. 

\vskip 5 truemm
\paragraph{Multiple scattering.}
The generalization of this simple result to $N$ scatterings is straightforward. After
integrating over the time of emission $t$ it is always possible to collect three pieces in
order to construct the effective radiation amplitude induced by $\vec{q}_i$ at time $t_i$.
Consider the three light-cone perturbation theory graphs of Fig.~5. 

\begin{figure}[h]
\label{fig5}
\centering
$M_{1}=
\begin{minipage}[b]{.25\textwidth}
\centering
\includegraphics[width=3.5cm]{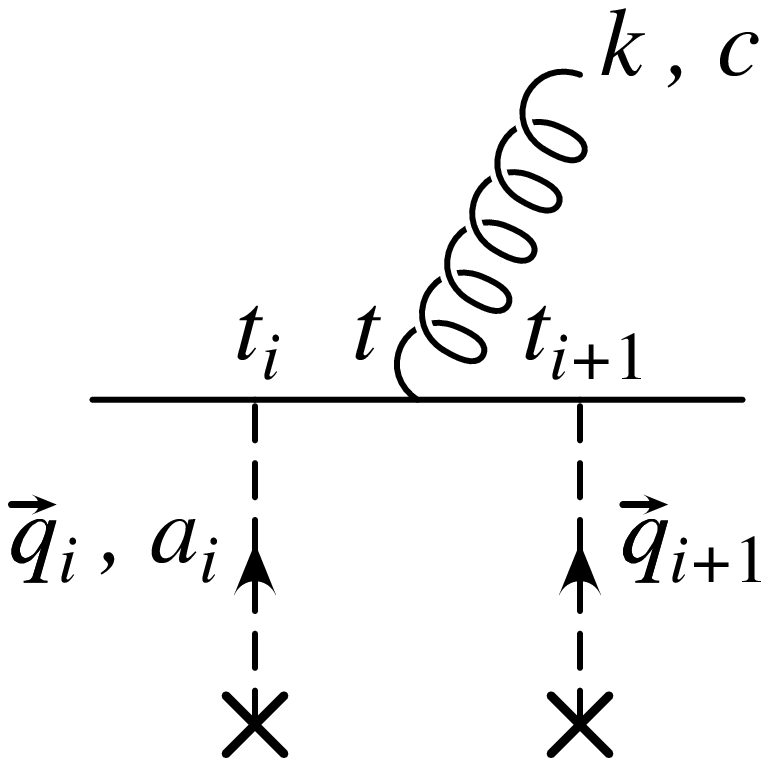}
\par\vspace{-30pt}
\end{minipage}
;M_{2}=
\begin{minipage}[b]{.25\textwidth}
\centering
\includegraphics[width=3.5cm]{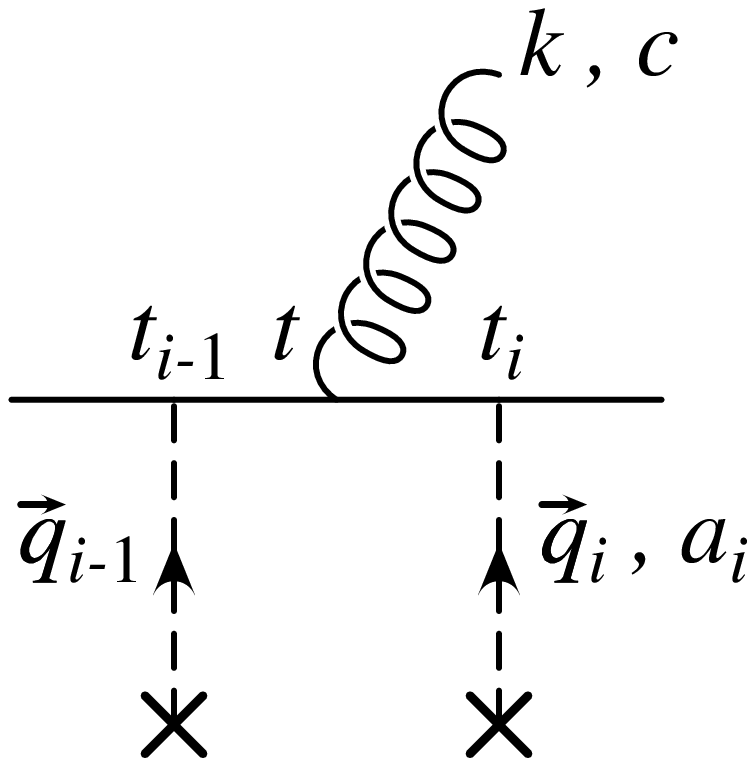}
\par\vspace{-30pt}
\end{minipage}
;M_{3}=
\begin{minipage}[b]{.25\textwidth}
\centering
\includegraphics[width=3.5cm]{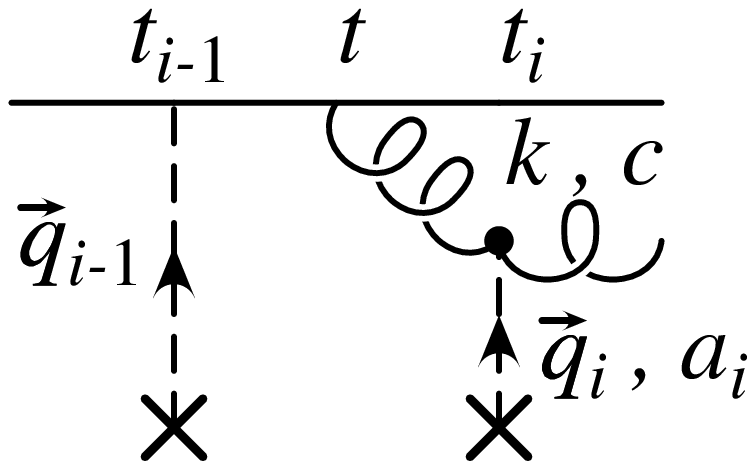}
\par\vspace{-30pt}
\end{minipage}
$
\vspace{30pt}
\caption{\it{Diagrams contributing to the effective emission amplitude induced by the transfer $\vec{q}_{i\bot}$.}}
\end{figure}

For each diagram, integrating over $t$ yields the
difference of two exponential phase factors (see Appendix A). Keeping only the one
depending on $t_i$, we collect three terms having the same phase, 
\bea
&&{\cal M}_1 \ \to \  - {\vec{k}_{\bot} \over k_{\bot}^2} e^{it_i {k_{\bot}^2 \over 2
\omega}} c a_i \ \ \ ,  \nn \\ 
&&{\cal M}_2 \ \to \  {\vec{k}_{\bot} \over k_{\bot}^2} e^{it_i {k_{\bot}^2 \over 2
\omega}} a_i c \ \ \ , \nn \\
&&{\cal M}_3 \ \to \  {\vec{k}_{\bot} - \vec{q}_{i\bot} \over (\vec{k} - \vec{q}_i)_{\bot}^2}
e^{i t_i{k_{\bot}^2 \over 2 \omega}} [c, a_i]  \ \ \ . \label{Effvertex} \eea
\noi The sum of these three terms gives the effective current

\beql{EffCurrent2} 
\vec{J}_{eff} (k, q_i) = \vec{J} (k, q_i) [c, a_i] = 
\begin{minipage}[c]{.3\textwidth}
\centering
\includegraphics[width=3.5cm]{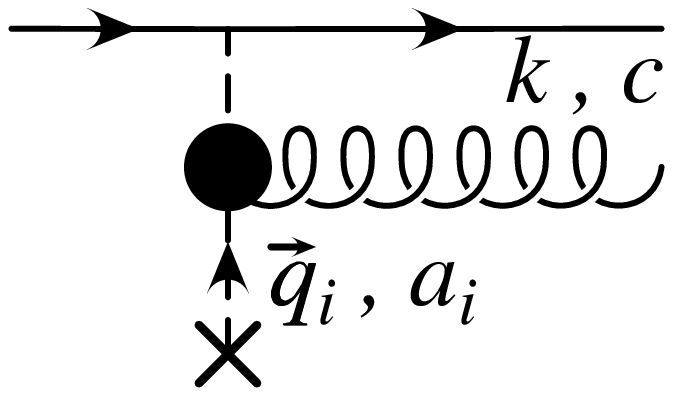}
\par\vspace{0pt}
\end{minipage}
\eeq

\noi as in (\ref{EffCurrent1}). 

Similarly to the case considered in (\ref{Mrad1}), the radiated gluon can rescatter on
centres $i\!+\!1,\ldots N$, so that the momentum $k$ and the colour factor have 
to be changed accordingly in (\ref{EffCurrent2}). 
For example, if the gluon emitted at centre $i$ rescatters on centre $j$ ($j > i$) 
the sum of the corresponding three terms results
in an expression analogous to (\ref{Effvertex}), with $k$ replaced by $(k - q_j)$, 
since $k$ labels the final real emitted gluon. In this case, one obtains 
\bminiG{EffCurrent3}
\vec{J}_{eff}(k - q_j, q_i) = \vec{J} (k - q_j , q_i) \ a_N ... a_{j+1} \ a_{j-1} \ldots
a_{i+1} [[c, a_j], a_i] \ . \label{EffCurrent31} 
\eeeq
This is diagrammatically shown below

\beeq
\vec{J}_{eff}(k-q_j,q_i)=
\begin{minipage}[c]{10cm}
{\includegraphics[height=4cm]{fig7.ps}}
\end{minipage} 
\emini

The associated phase is shifted according to
\beq\label{phaseshift} 
\exp\left\{it_i{k_{\bot}^2 \over 2\omega}\right\} 
\Longrightarrow \exp\left\{it_i {(\vec{k} - \vec{q}_j)_{\bot}^2 \over 2
\omega} + it_j {k_{\bot}^2 - (\vec{k}-\vec{q}_j)_{\bot}^2 \over 2 \omega}\right\} \ .
\eeq
In the total radiation amplitude, we should include, for centre $i$, the $2^{N-i}$
possibilities (labelled by $i_r$) for the quark-gluon system to rescatter on the
remaining centres. For $r = 1$ to $r = 2^{N-i}$, the associated phase gets modified each
time the gluon rescatters (the phase is unchanged by quark rescattering). Thus we write
\beq \label{Mrad} 
M_{rad} \propto \sum_{i=1}^N \sum_{r=1}^{2^{N-i}} \vec{J}_{eff} (k_{i_r}, q_i )
e^{i \varphi_{i_r}} \> ,
\eeq
where the colour structure is included in $\vec{J}_{eff}$.

\subsection{Expression for the radiation spectrum induced by $N$ scatterings}

As for a single scattering, we square the radiation amplitude $M_{rad}$ given in
(\ref{Mrad}) and normalize by the multiple elastic scattering cross section to get the
radiation spectrum induced by $N$ scatterings
\beq\label{SpectrumiN}
 \omega {dI^{(N)} \over d \omega} = {\alpha_s \over \pi^2} \int d^2
\vec{k}_{\bot} \left < \sum_{i=1}^N \sum_{j=1}^N {\vec{J}^i_{eff} \cdot 
\vec{J}^{j\dagger}_{eff}
\over N_c \ C_F^{\ N}} \ e^{i(\varphi_i - \varphi_j)}  \right > \> . 
\eeq 
This expression deserves some comments.

\begin{figure}[h]
\label{fig8}
\centering
\begin{minipage}[c]{\textwidth}
\centering
\includegraphics[height=4cm]{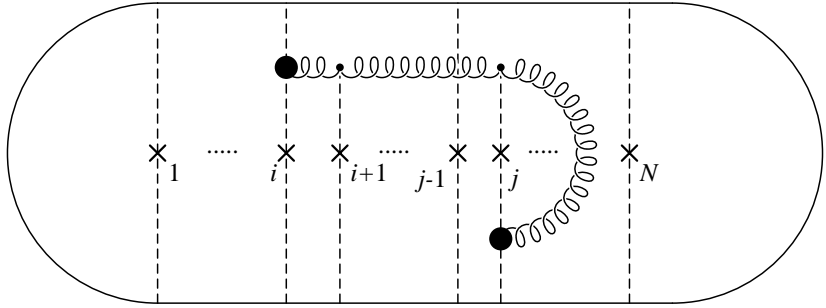}
\par\vspace{0pt}
\end{minipage}

\caption{\it{Interference term $\vec{J}^i_{eff} \cdot (\vec{J}^j_{eff})^\dagger$ in the form of a connected graph.}}
\end{figure}

\ben
\item
In $\vec{J}^i_{eff}$, the index $i$ refers to the centre which induces the effective
emission current. For a simple calculation of colour factors, it is convenient to represent
interference terms in the form of connected diagrams, where the ``conjugate amplitude''
$(\vec{J}^j_{eff})^\dagger$ 
appears in the lower part of the diagram (Fig.~6). 

\item 
The colour factor in the denominator of (\ref{SpectrumiN}) corresponds to the
normalization to the elastic scattering cross section, depicted in Fig.~7.
 
\begin{figure}[h]
\label{fig9}
\centering
\begin{minipage}[c]{.25\textwidth}
\end{minipage}
\begin{minipage}[c]{.5\textwidth}
\includegraphics[width=\textwidth]{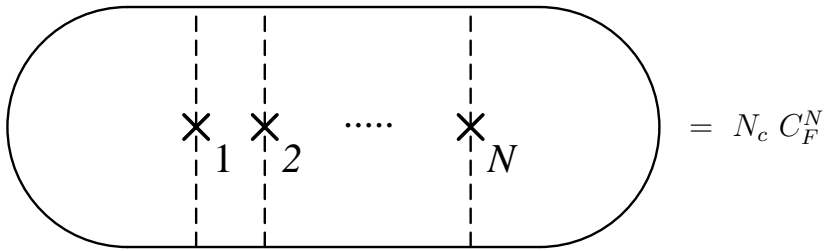}
\end{minipage}
\begin{minipage}[c]{.25\textwidth}
$\ \ =\  N_c \ C_F^N$
\end{minipage}

\caption{\it{Colour factor associated with the multiple elastic scattering cross section.}}
\end{figure}
This is easily calculated from the rules given in Appendix B. 

\item 
The sum over all possible gluon rescatterings is implicit in (\ref{SpectrumiN}) 
(the sum over $r$ in (\ref{Mrad})). We should take into account 
all possible ways the quark-gluon system has to rescatter on centres $n$,
in particular, for $n \geq j +1$. 
However this part of the diagram describes the multiple scattering
of the produced quark-gluon system after centre $j$, 
which has no influence on the energy radiation spectrum we are interested in. 

\item  
Between centres $i$ and $j$, it matters whether it is the quark or the gluon which
absorbs the transverse momentum $\vec{q}_n$, because it changes the relative phase
$(\varphi_i - \varphi_j)$. To simplify our derivation, we will first consider the large
$N_c$ limit, where all non-planar diagrams may be dropped, 
which corresponds to neglecting quark scatterings on centres 
$n$ for $i < n < j$. 

\item
Finally, the brackets in (\ref{SpectrumiN}) denote the average over 
momentum transfers and longitudinal coordinates of the scattering centres. 
As in \cite{BDMPS} we have
\beq
\left < (\ \ ... \ \ ) \right > \Leftrightarrow \int \prod_{\ell = 1}^{N-1} {d \Delta_{\ell}
\over \lambda} \exp \left ( - {\Delta_{\ell} \over \lambda} \right ) \cdot \int
\prod_{i=1}^N d^2 \vec{q}_{i_{\bot}} \ V(q_{i_{\bot}}^2) (\ \ ... \ \  ) \
,\label{[Average]}  
\eeq
where $\Delta_{\ell} = z_{\ell + 1} - z_{\ell}$ and $V(q_{\bot}^2)$ is the normalized
cross section for elastic scattering, given by (\ref{Potential}) in the case of Coulomb
potentials. 
\een

The products $\vec{J}^i_{eff} \cdot \vec{J}^{j\dagger}_{eff}$ 
in (\ref{SpectrumiN}) include a colour sum as indicated in (\ref{fig10}).
We may rewrite (\ref{SpectrumiN}) as
\bminiG{q-def2}
 \omega {dI^{(N)} \over d \omega} = {\alpha_s \over \pi^2}  \int
d^2\vec{k}_{\bot} \left < 2 {\rm Re} \sum_{i=1}^N \sum_{j=i+1}^N {\vec{J}^i_{eff} \cdot
\vec{J}^{j\dagger}_{eff} \over N_c C_F^{j-i+1}} \ e^{i(\varphi_{_i} - \varphi_{_j})} 
+ \sum_{i=1}^N |\vec{J}^i_{eff}|^2 \right >  \  . \label{Spectrum1}  
\eeeq
As in the case of QED \cite{BDMPS}, we have the equivalent expression
\beeq 
 \omega {dI^{(N)} \over d\omega}  = {\alpha_s \over \pi^2} \int d^2
\vec{k}_{\bot} \left < 2 \ {\rm Re} \sum_{i=1}^N \sum_{j=i+1}^N  
{\vec{J}^i_{eff} \cdot \vec{J}^{j\dagger}_{eff} \over N_c \ C_F^{j-i+1}} 
\left ( e^{i (\varphi_{_i} - \varphi_{_j})} - 1\right ) 
+  \left | \sum_{i=1}^N \vec{J}^i_{eff} \right |^2 \right > \ .  
\label{Spectrum2}  
\emini
In the large-$N_c$ limit 
$\vec{J}^i_{eff} \cdot \vec{J}^{j\dagger}_{eff}$ is given by
two sets of planar diagrams denoted by $Y$ and $H$ and shown in (\ref{fig10}). 
\bminiG{fig10}
Y & = & 
\begin{minipage}[c]{.6\textwidth}
\includegraphics[width=\textwidth]{fig10a.ps}
\end{minipage} 
 \\[5mm]
H & = & 
\begin{minipage}[c]{.6\textwidth}
\includegraphics[width=\textwidth]{fig10b.ps}
\end{minipage}
\emini

\noi 
The second term of (\ref{Spectrum2}) is the so-called factorization term, which
corresponds to the limit of vanishing phases. 
In this limit, all emission amplitudes from the internal lines vanish 
(see Appendix A). 
Two cases have to be distinguished. 

\bit
\item
If the incident quark is produced at a time $t_0 = - \infty$, we see from Appendix A
that only emission amplitudes from initial and final lines remain. 
The factorization term contribution is then equivalent to the contribution induced 
by a single scattering of momentum transfer 
$\vec{q}_{\bot_{tot}} = \sum\limits_{i=1}^N \vec{q}_i$, 
and thus has a weak logarithmic medium dependence, as in the QED case~\cite{BDMPS}.

\item 
In a realistic situation where the incident quark is produced, through 
a hard scattering, at a time $t_0 = 0$, only emission from the final line remains
(see the table of Appendix A). In this case the factorization term has no medium 
dependence at all, so that the medium induced spectrum is exactly given 
by the first term of (\ref{Spectrum2}). It should be directly accessible 
experimentally by comparing hard scattering on a nucleus with that on a proton. 
\eit

We show in Appendix C that after dropping the medium-independent 
factorization term, (\ref{Spectrum2}) leads to the following medium-induced 
radiation spectrum in the large-$N_c$ limit  
\beq\label{LPM} 
\omega {dI^{(N)} \over d \omega}  = {3 \over 2} N_c {\alpha_s \over \pi^2} 
\int d^2 \vec{U} \left < 2 {\rm Re} \sum_{i=1}^N \sum_{j=i+1}^N \vec{J}_i 
\cdot \vec{J}_j \left [\exp \left \{ i \kappa \sum_{\ell = i}^{j-1} U_{\ell}^2 
{\Delta_{\ell} \over \lambda} \right \} - 1 \right ]  \right > \ , 
\eeq
where the $\vec{U}_i$'s are the transverse momenta of the gluon 
expressed in units of $\mu$,
\bea
&&\vec{U}_i = \vec{U}_{i-1} - \vec{Q}_i \nn \\
&&\vec{U} = {\vec{k}_{\bot} \over \mu} \ ; \ \vec{Q}_i = {\vec{q}_i \over \mu} \
,\label{Rescaling} 
\eea
and $\vec{J}_i$ is the rescaled emission current
\beq\label{kinCurrent}
\vec{J}_i = {\vec{U}_i \over U_i^2} - {\vec{U}_{i-1} \over U_{i-1}^2} \ . 
\eeq
The dimensionless parameter $\kappa$ is
\beq\label{kappa}
\kappa = {\lambda \mu^2 \over 2} {1 \over \omega} \ . 
\eeq
The radiation spectrum for the infinite medium QED case was given in 
\cite{BDMPS} for $\kappa\ll 1$.
We observe that the expression (\ref{LPM}) has the same form 
as in QED \cite{BDMPS}, with the replacements
\bminiG{q-def3}
\alpha \rightarrow {3 \over 2} N_c \ \alpha_s \qquad ; \qquad
{\omega \over E^2} \rightarrow {1 \over \omega} \ \ \ , \label{QEDQCD}
\eeeq
\beeq
\hbox{photon ``angle''}\  \vec{U}_i \rightarrow \hbox{gluon ``transverse momentum''} \
\vec{U}_i  \ \ \ . \label{anglemomentum}
\emini
This analogy allows us to give directly the result for the infinite 
QCD medium. 
Thus,
\beq
\omega {dI \over d \omega dz} = {3 \alpha_s \over 2 \pi} \ {N_c \over \lambda}
\ \sqrt{\kappa \ \ln {1 \over \kappa}} \ \ \ .
 \label{36} 
\eeq
Note that the radiation density is obtained by normalizing (\ref{LPM}) 
to the distance $N\lambda$, where $\lambda$ is the mean free path of the incident 
quark \cite{BDPS}. 

The changes necessary to include all $\displaystyle{{1 \over N_c}}$ corrections 
as well as the case of an arbitrary incident parton of colour represention $R$ 
have been worked out in \cite{BDPS}. These changes are 
\bminiG{q-def4}\label{finiteNca} 
\kappa &\to& \widetilde{\kappa} = {\widetilde{\lambda} \mu^2 \over 2} {1 \over
\omega} \ ; \qquad \widetilde{\lambda} = {2C_F \lambda \over N_c} = 2 \lambda_g \>, \\
\label{finiteNcb}
{N_c \over \lambda} &\to& {N_c \over \lambda_R} = {C_R \over \lambda_g} \ . 
\emini
This leads to the general formula
\beq
\omega {dI \over d \omega dz} = {3 \alpha_s \over 2 \pi} {C_R \over \lambda_g}
\sqrt{\widetilde{\kappa} \ln {1 \over \widetilde{\kappa}}} = {3 \alpha_s \over 2 \pi}
\ C_R \sqrt{{\mu^2 \over \lambda_g \omega} \ln \left ( {\omega \over \mu^2 \lambda_g}
\right ) } \ , \label{QCDBDMPS} 
\eeq 
where $\lambda_g$ is the gluon mean free path. 

We note that apart from the overall normalization, proportional to the squared colour
charge $C_R$, there is no dependence of the induced radiation spectrum on the nature 
of the initial parton.

We will now study, for a finite medium, the dependence of the radiation
spectrum on the medium size $L = N\lambda$. 
We present first a heuristic discussion of 
finite-size effects, and will use (\ref{LPM}) as a starting point in section~\ref{sec:nom4}. 

 
\mysection{Heuristic discussion of the energy loss in finite length media}
\label{sec:nom3}
When a very energetic parton of energy $E$ is propagating through a medium of finite
length\break \noindent $L = N \lambda$ the gluon radiation spectrum shows characteristic
features depending on the gluon energy $\omega$. For discussing the radiation density
$\omega \displaystyle{{dI / d\omega}}$ three different regimes may be distinguished
\citd{BDPS}{BDMPS} : the Bethe-Heitler (BH) regime with small gluon energies, 
the coherent regime (LPM) for intermediate $\omega$, and the highest 
energy regime corresponding to the factorization limit. 
The coherent regime corresponds to the condition (cf. (4.19a) in \cite{BDMPS}) 
\beq
{1\over N^2} \ll \kappa \ll 1 \ \ \ , \label{3.1}
\eeq
with $\kappa$ given in (\ref{kappa}) for the QCD case. 
Thus, for the finite media
under consideration a reasonably large number of scatterings 
will be assumed, $N \gg 1$. 

For the following qualitative derivations we neglect logarithmic factors. Thus we ignore
numerical factors of order 1, and do not distinguish between propagating quarks and
gluons. However we explicitly keep the parameters representing the medium.

In terms of the gluon energy the condition (\ref{3.1}) is
\beq \label{3.2} 
\omega_{BH} \sim \lambda \mu^2 \ \ll \ \omega \ \ll \ \omega_{fact} \sim {\mu^2L^2
\over \lambda} \leq E \> .
\eeq
Obviously, $\omega_{fact} \leq E$ only holds when $L$ is less than the critical length,
\beq \label{3.3}
L \leq L_{cr} = \sqrt{\frac{\lambda E}{\mu^2}} \> .
\eeq
\noi We note that the case $\omega_{fact} \ll E$ is consistent with 
the soft gluon approximation for the induced spectrum. 

The radiation spectrum per unit length behaves in the $E \to \infty$
limit as 
\beq
\omega {dI \over d \omega dz} \simeq \left \{ \begin{array} {ll}
\displaystyle{{\alpha_s \over \lambda}} &\qquad \omega < \omega_{BH} \\
\displaystyle{{\alpha_s \over \lambda}} \sqrt{\displaystyle{{\lambda \mu^2
\over \omega}}} &\qquad \omega_{BH} < \omega < \omega_{fact} \\
& \\ 
\displaystyle{{\alpha_s \over L}} &\qquad \omega_{fact} < \omega < E \\ 
\end{array} \right . \label{3.4}
\eeq
\noi for a finite length $L \leq L_{cr}$. These main features are illustrated 
schematically in 
Fig.~8. 
In the BH regime the radiation is due to $N = L/\lambda$ 
incoherent scatterings, whereas in the factorization regime the medium behaves as 
one single scattering centre. 
In the LPM regime $1/\sqrt{\kappa}$ elementary centres act as a single
scattering centre.

In order to obtain the total energy loss $\Delta E$ we integrate the spectrum 
(\ref{3.4}) over $\omega$ and $z$, with $0 \leq \omega \leq E$ and $0 \leq z \leq L$. 
In addition to medium-independent contribution to the energy loss proportional 
to $\alpha_s E$ (a factorization contribution), 
we find the induced loss
\beq\label{3.5} 
\Delta E(L) \sim \alpha_s {\mu^2 L^2 \over \lambda} \left ( 1 + O \left (
\displaystyle{{1 \over N}} \right ) \right) .
\eeq 
It has been already pointed out in \cite{AM}, that the total energy
loss increases quadra\-ti\-cally with the length $L$, and is independent of the parton
energy $E$ in the high energy limit. This interesting case is investigated in
more detail in what follows. 

Here we conclude this heuristic discussion by  considering the case of finite
$L$, but with $L > L_{cr}$. This situation occurs for parton energies $E \leq E_{cr} =
L^2 \mu^2/\lambda$ (see (\ref{3.3})). By extending the coherent soft $\omega$-spectrum
(\ref{3.4}) up to $\omega \sim E$, the total induced loss is
\beq \label{3.6}
\Delta E(L)\>\simeq\>\alpha_s \sqrt{{\mu^2 E \over\lambda}}\, L \>=\>
\alpha_s \sqrt{E \ E_{cr}} \ \ \ , 
\eeq
which is $E$-dependent and linear in $L$. The results given in (\ref{3.5}) and
(\ref{3.6}) are schematically summarized in Fig.~9, 
where the $E$ dependence of 
$\Delta E$ is plotted for fixed $L$, and in Fig.~10, 
where $E$ is fixed and the $L$-dependence is shown.

The loss given by (\ref{3.6}) is also relevant for an infinite medium, 
$L \to \infty$. 
An $E$-dependent loss per unit length of propagating partons is found \cite{BDPS},
\beq
- {dE \over dz} \> \simeq \> \alpha_s \sqrt{{\mu^2 E\over \lambda}} \>. 
\eeq

\vskip 1 truecm
\begin{figure}[h]
\label{fig11}
\centering
\includegraphics[height=6cm]{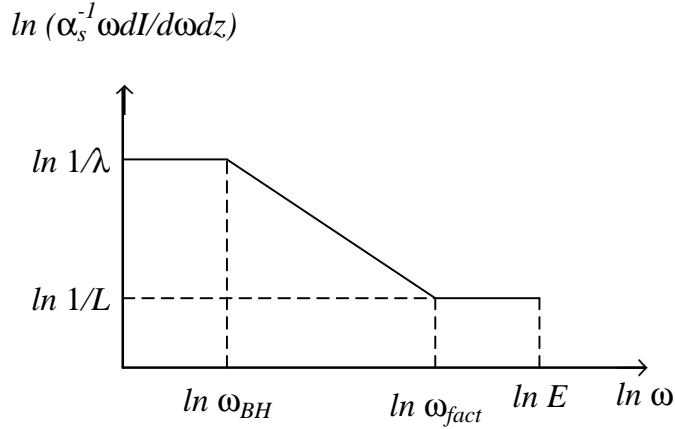}
\caption{\it{Radiation density (\ref{3.4}) for a medium of finite length $L < L_{cr}$.}}
\end{figure}


\begin{figure}[h]

\label{fig12et13}
\begin{minipage}[b]{.45\linewidth}
\centering
\includegraphics[height=5cm]{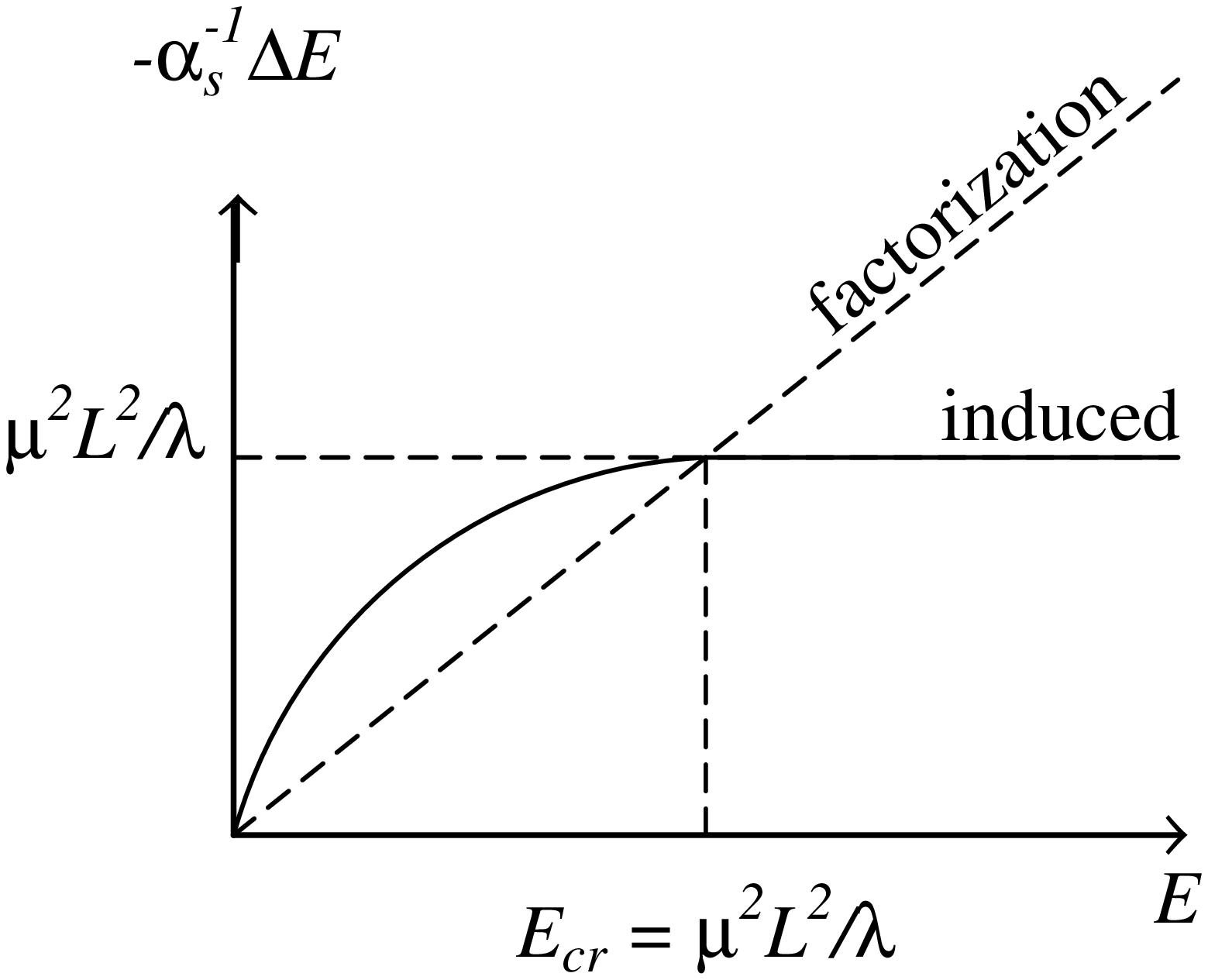}
\caption{\it{Total induced energy loss as a function of the parton energy $E$.}}
\label{fig12}
\end{minipage}%
\hspace{.05\linewidth}
\begin{minipage}[b]{.45\linewidth}
\centering
\includegraphics[height=5cm]{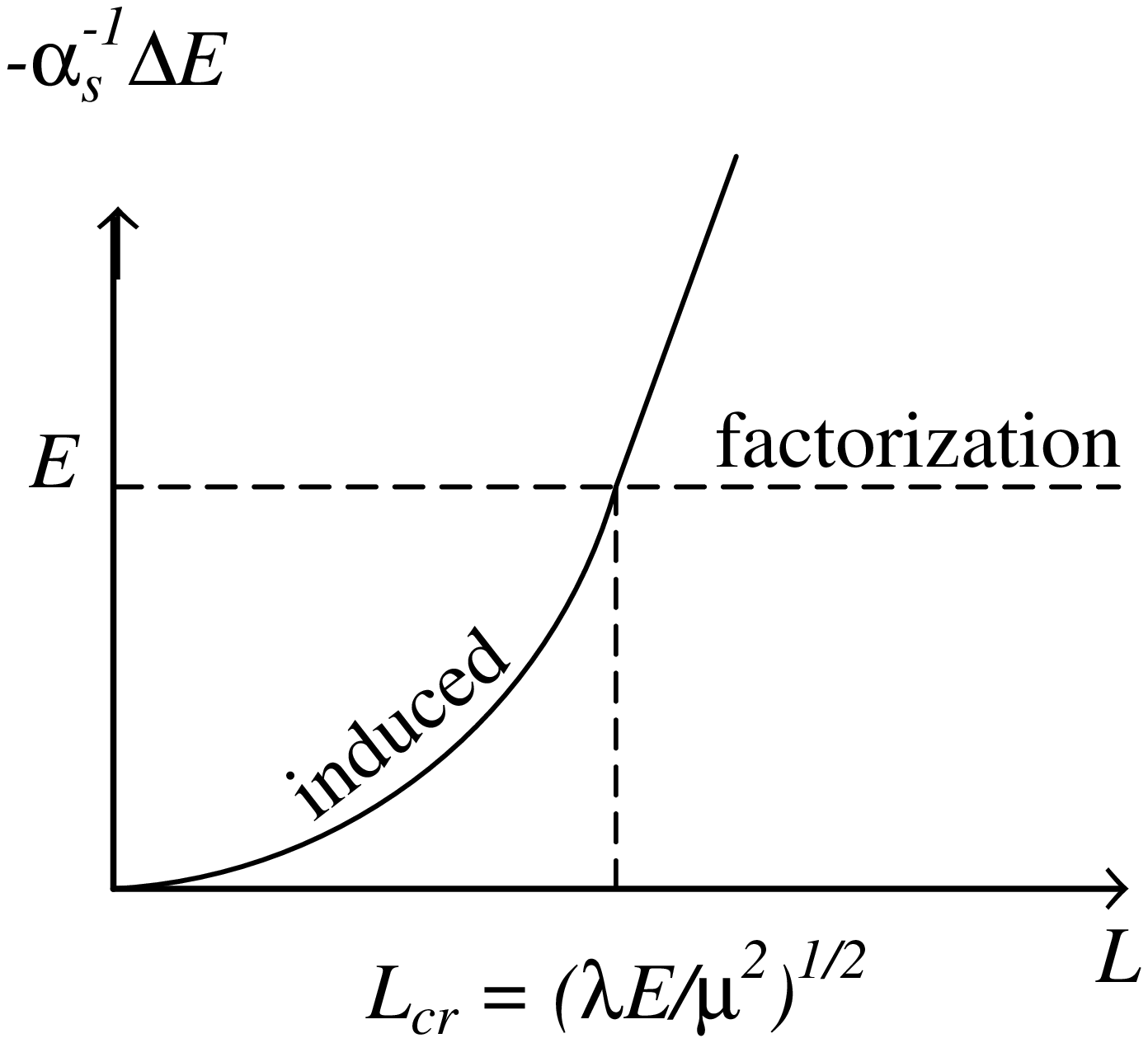}
\caption{\it{Total induced energy loss as a function of the medium size $L$.}}\label{fig13}
\end{minipage}%

\end{figure}

\mysection{General equation for the induced radiation spectrum}
\label{sec:nom4}
In this section we shall derive the general equations which govern the induced radiation
spectrum for finite length materials. These equations generalize 
(4.34) and (4.40) of BDMPS. Our starting point, for the sake of simplicity, is the
large-$N_c$ formula (\ref{LPM}). We divide by $L$, the length of the material, 
and we allow the sum over scatterings to be arbitrary in number giving
\beq\label{4.1} 
{\omega\> dI \over d\omega dz} = {3\alpha_s N_c \over 2L \pi^2} \int_0^L d\Delta 
\int d^2 \vec{U} \left < 2 {\rm Re} \sum_{i=1}^{\infty} \sum_{j = i+1}^{\infty} 
\vec{J}_i \cdot \vec{J}_j \left [ \exp \left \{ i \kappa \sum_{\ell = i}^{j-1} 
U_{\ell}^2 {\Delta_{\ell} \over \lambda} \right \} - 1 \right ] 
\delta \left ( \Delta - \sum_{m=1}^{\infty} \Delta_m \right ) \right> , 
\eeq 
where $\Delta$ is the distance between the first
and last scatterings. Expressing the sum over $i$ as an integral of the position in the
medium of the gluon emission vertex $\vec{J}_i$ and neglecting initial and final state
scattering one arrives at 
\[
\omega {dI \over d \omega dz} = {3 \alpha_s N_c \over 2L \pi^2} \int_0^L d \Delta
\int_0^{L- \Delta} {dz_1 \over \lambda} \int d^2 \vec{U}  
\]
\beq\label{spec2}
\cdot \left < 2 {\rm Re}
\sum_{n=0}^{\infty} \vec{J}_1 \cdot \vec{J}_{n+2} \left[\exp\left\{ i\kappa 
\sum_{\ell = 1}^{n+1} U_{\ell}^2 {\Delta_{\ell} \over \lambda} \right \} - 1 \right ] 
\delta  \left (\Delta - \sum_{m=1}^{n+1} \Delta_m \right ) \right > , 
\eeq
where now $\Delta$ is the distance between the emission vertices $\vec {J}_1$ and
$\vec{J}_{n+2}$. Although the sum in (\ref{spec2}) formally goes from $n = 0$ to 
$n = \infty$
we expect the typical number of scatterings to be $L/\lambda$. 
It is convenient to scale all distances by the mean free path $\lambda$. 
To that end let 
\[ 
t = {\Delta \over \lambda} \quad , \quad t_{\ell} =
{\Delta_{\ell} \over \lambda} \quad , \quad T = {L \over \lambda} \> . 
\] 
Then
\[
\omega {dI \over d \omega dz} = {3 \alpha_s N_c \over 2 \lambda \pi} \int_0^T dt \> 
(1 - {t\over T}) \int {d^2\vec{U} \over \pi} 2{\rm Re} \sum_{n=0}^{\infty} 
\int \prod_{\ell =
1}^{n+2} d^2 \vec{Q}_{\ell} V(Q_{\ell}^2) \vec{J}_1 \cdot \vec{J}_{n+2}  
\]
\beq\label{spec3}
\cdot \int \prod_{r=1}^{n+1}
e^{-t_r} dt_r \left [ \exp \left \{ i \kappa \sum_{s=1}^{n+1} U_s^2 t_s \right \} - 1
\right ] \delta \left ( t - \sum_{m=1}^{n+1} t_m \right ) ,
\eeq
where we have done the integral over $z_1$. In (\ref{spec3}) the dependence on
$\vec{Q}_1$ and $\vec{Q}_{n+2}$ is contained only in the product of currents
\beq
\vec{J}_1 \cdot \vec{J}_{n+2} = \left ( {\vec{U}_1 \over U_1^2} - {\vec{U}_1 + \vec{Q}_1
\over ( \vec{U}_1 + \vec{Q}_1 )^2} \right ) \cdot \left ( {\vec{U}_{n+1} - \vec{Q}_{n+2}
\over ( \vec{U}_{n+1} - \vec{Q}_{n+2})^2} - {\vec{U}_{n+1} \over U_{n+1}^2} \right ) ,  
\label{4.4}
\eeq 
and in $V(Q_1^2)$ and $V(Q_{n+2}^2)$. 
As in BDMPS we integrate first over $\vec{Q}_1$ and $\vec{Q}_{n+2}$ holding 
$\vec{U}_1$, $\vec{U}_2, \ldots \vec{U}_{n+1}$ fixed. 
Defining an averaged elementary current
\beq\label{4.5}
\vec{f}_0 (\vec{U}_1) \equiv \int d^2 \vec{Q}_1 V(Q_1^2) \vec{J}_1 
= \pi {\vec{U}_1 \over U_1^2} \int_{U_1^2}^{\infty} dQ^2V(Q^2) \> , 
\eeq
we note that
\beq\label{4.6} 
\vec{f}_0(\vec{U}_{n+1}) = - \int d^2 \vec{Q}_{n+2} \ V(Q_{n+2}^2) \ 
\vec{J}_{n+2} \>. 
\eeq
The spectrum can be written as
\beq \label{4.7}
\left . \omega {dI \over d \omega dz} = {3 \alpha_s N_c \over 2 \pi \lambda} 2 {\rm Re}
\int_0^T dt \ (1 - {t \over T} ) \int {d^2 \vec{U}_1 \over \pi} \vec{f}_0(\vec{U}_1) 
\cdot \vec{f}(\vec{U}_1, t) \right |_{\kappa}^{\kappa = 0} \> ,
\eeq 
where
\[
\vec{f} (\vec{U}_1, t) = \vec{f}_0 (\vec{U}_1) e^{-t(1 - i \kappa U_1^2)} + e^{-t(1 - i
\kappa U_1^2)} \sum_{n=1}^{\infty} \prod_{\ell = 2}^{n+1} \int d^2 \vec{Q}_{\ell}
V(Q_{\ell}^2) \]
\beq \label{f} 
\cdot \prod_{r=2}^{n+1} \int dt_r \ e^{-i\kappa t_r (U_1^2 - U_r^2)} \> 
\Theta  \left ( t -\sum_{m=2}^{n+1} t_m \right ) \vec{f}_0(\vec{U}_{n+1}) \> .
\eeq
The first term on the right-hand side of (\ref{f}) is the $j=i\!+\!1$ (nearest
neighbours, $n\!=\!0$) term of (\ref{LPM}). 
The rest of the terms in (\ref{f}) correspond to arbitrary numbers of gluon
rescatterings between the currents 1 and $n\!+\!2$.

Equation (\ref{f}) has a Bethe-Salpeter structure so it is straightforward to check that
$\vec{f}(\vec{U}, t)$ satisfies the natural integral equation
\beq\label{IntegralEq}
\vec{f}(\vec{U}, t) = \vec{f}(\vec{U}, 0) e^{-t(1 - i \kappa U^2)} + \int d^2 \vec{Q}
\ V(Q^2) \int_0^t dt' \ e^{-(t-t')(1 - i \kappa U^2)} \vec{f} (\vec{U} - \vec{Q}, t') ,
\eeq 
where
\beq \label{4.10}
\vec{f}(\vec{U}, 0) \equiv \vec{f}_0(\vec{U}) \> ,
\eeq
and where we have used
\beq\label{4.11}
\vec{U}_{\ell} = \vec{U}_{\ell - 1} - \vec{Q}_{\ell} \> . 
\eeq
It is convenient to eliminate the $t$-integration on the right-hand side of
(\ref{IntegralEq}). This is achieved by taking a $t$-derivative 
which gives  
\beq\label{DiffEq}
{\partial \over \partial t} \vec{f}(\vec{U}, t) = - (1 - i \kappa U^2) \> \vec{f} 
(\vec{U} ,t) + \int d^2 \vec{Q} \ V(Q^2) \ \vec{f} (\vec{U} - \vec{Q}, t) \> . 
\eeq
Although (\ref{DiffEq}) has been derived for an incident quark 
in the large-$N_c$ limit, essentially the same
equation holds for partons of colour representation $R$ even without taking 
the large-$N_c$ limit. In the general case
\beeq 
\vec{f}(\vec{U},t_R) &=& {N_c \over 2C_R} \vec{f}_0(\vec{U}) 
\> e^{-t_R(1 - i \kappa_R U^2)} + {N_c \over 2C_R} \int\! d^2 \vec{Q} \, V(Q^2) 
\int_0^{t_R} \! dt'_R \ e^{-(t_R - t'_R) (1 - i \kappa_R U^2)} \vec{f} 
(\vec{U} \!-\! \vec{Q}, t'_R) \nonumber \\
&+& \left ( 1 - {N_c \over 2 C_R} \right ) \int^{t_R}_0 dt'_R \ e^{-(t_R - t'_R) (1 - i
\kappa_R U^2)} \vec{f} (\vec{U}, t'_R) \label{IntegralEqNc}
\eeeq
replaces (\ref{IntegralEq}) with $t_R=z/\lambda_R$, $\kappa_R=\mu^2\lambda_R/2\omega$ 
and where $\lambda_R = \lambda C_F/C_R$. 
The second term on the right-hand side of (\ref{IntegralEqNc}) corresponds 
to the emitted gluon rescattering which is dominant for incident quarks
in the large $N_c$ limit, 
and thus corresponds to the second term on the right-hand side of (\ref{IntegralEq}). 
The third term on the right-hand side of (\ref{IntegralEqNc}) is necessary to generate 
the non-planar graphs corresponding to the incident parton rescattering in the medium 
as discussed in BDPS. 
Indeed, as already discussed in section \ref{sec:nom2}, there is no
additional phase factor associated with the incident parton rescattering. 
As a result, this term is $Q$-independent and we can use
$\int d^2\vec{Q}\, V(Q^2)=1$.
 
The colour factors appearing in (\ref{IntegralEqNc}) may be easily obtained from 
Appendix B. 
Taking a time derivative on both sides of (\ref{IntegralEqNc}), and defining 
\beq 
\tau = {N_c \over 2C_R} t_R \quad ,
\quad \tau_0 = {N_c \over 2 C_R} T_R \>  , \label{4.14}  
\eeq 
\beq 
\widetilde{\kappa} = {2C_R \over N_c} \kappa \> , \label{4.15}  
\eeq
we find
\beq
{\partial \over \partial \tau} \vec{f}(\vec{U}, \tau ) = - (1 - i
\widetilde{\kappa} U^2) \vec{f} (\vec{U}, \tau ) + \int d^2 \vec{Q} \ V(Q^2) \ \vec{f}
(\vec{U} - \vec{Q} , \tau ) \> , \label{DiffEqNc} 
\eeq 
which equation is identical in form to (\ref{DiffEq}). 
It is remarkable that the equation for the spectrum keeps the simple form
\beq
\left . \omega {dI \over d \omega dz} = {3 \alpha_s C_R \over 2 \pi \lambda_g} 2 {\rm
Re} \int_0^{\tau_0} d\tau \left ( 1 - {\tau \over \tau_0} \right ) \int {d^2
\vec{U}_1 \over \pi} \vec{f}_0 (\vec{U}_1) \cdot \vec{f}(\vec{U}_1, \tau )
\right |_{\tilde{\kappa}}^{\tilde{\kappa} =0} \ \ \ , \label{specNc} 
\eeq 
provided we take with (\ref{DiffEqNc}) the initial condition 
\beq 
\vec{f}(\vec{U}, 0) = \vec{f}_0
(\vec{U}) \ \ \ , \label{4.18} 
\eeq
instead of the initial condition implied in (\ref{IntegralEqNc}). 

We can make contact with the formalism of BDMPS for infinite length matter by taking
$\tau_0 = \infty$ in (\ref{specNc}). Identifying
\beq
\vec{f}(\vec{U}) = \int_0^{\infty} d \tau \ \vec{f}(\vec{U}, \tau ) \ \ \ , \label{fvec}
\eeq
\noi we see that (\ref{specNc}) agrees with (4.28) of BDMPS when account is taken of the
replacement $\alpha \to \displaystyle{{3 \over 2}} \alpha_s N_c$ necessary to go 
from QED to QCD. Integrating (\ref{DiffEqNc}) over $\tau$ between $\tau = 0$ and 
$\tau = \infty$ gives
\beq
(1 - i \widetilde{\kappa}U^2) \ \vec{f}(\vec{U}) = \vec{f}_0 (\vec{U}) + \int d^2
\vec{Q} \ V(Q^2) \ \vec{f}(\vec{U} - \vec{Q}) \ \ \ , \label{4.20} \eeq
\noi which again reproduces the BDMPS result given in (4.30). \par

Just as (4.20) was solved, for small $\widetilde{\kappa}$, by going to impact 
parameter space, 
it is convenient to recast (\ref{DiffEqNc}) by defining
\beq
\widetilde{\vec{f_{\ }}} (\vec{B}, \tau ) = \int d^2 \vec{U} \ e^{-i \vec{B}\cdot
\vec{U}} \ \vec{f} (\vec{U}, \tau) \ \ \ , \label{ftilde} \eeq
\beq
\widetilde{V}(B^2) = \int d^2 \vec{Q} \ e^{-i \vec{B}\cdot \vec{Q}} V(Q^2) \ \ \
, \label{4.22} \eeq
\noi which leads to
\beq
{\partial \over \partial \tau} \widetilde{\vec{f_{\ }}}(\vec{B}, \tau ) = \left [ - i
\widetilde{\kappa} \nabla_B^2 - (1 - \widetilde{V}(B^2)) \right ]
\widetilde{\vec{f_{\ }}}(\vec{B}, \tau ) \ \ \ . \label{Diffusion1} 
\eeq 
We can also use (\ref{ftilde}) to rewrite the spectrum 
(\ref{specNc}) as an integral over impact
parameter with the result 
\beq
\omega \left . {dI \over d \omega dz} = {3 \alpha_s C_R \over 2 \pi^2 \lambda_g} 
2\, {\rm Re} 
\left\{ i \int_0^{\tau_0} d\tau \left ( 1 - {\tau \over \tau_0} \right ) \int {d^2\vec{B} \over
2 \pi} \ {1 - \widetilde{V}(B^2) \over B^2} \vec{B} \cdot \widetilde{\vec{f_{\ }}}(\vec{B},
\tau ) \right |_{\tilde{\kappa}}^{\tilde{\kappa} = 0}\right\}, \label{Bspec} 
\eeq 
the finite length generalization of (4.40) of BDMPS. 
In arriving at (\ref{Bspec}) we have used 
\beq 
\widetilde{\vec{f_{\ }}}_0(\vec{B}) = - 2 \pi i {\vec{B} \over B^2}
(1 -\widetilde{V}(B^2)) \> .
 \label{4.25} 
\eeq
Equations (\ref{Diffusion1}) and (\ref{Bspec}), along with the initial condition
$\widetilde{\vec{f_{\ }}}(\vec{B}, 0) = \widetilde{\vec{f_{\ }}}_0(\vec{B})$ are the general equations
governing the induced radiation spectrum for finite length matter in the context of the
Gyulassy-Wang model of a hot QCD plasma.

\mysection{Solution for the radiation spectrum in the small $\kappa$ limit}
\label{sec:nom5}
In this section we give an approximate solution to (\ref{Diffusion1}) and (\ref{Bspec}) valid
in the small $\kappa$ limit. We restrict our evaluation of (\ref{Bspec}) to $\omega$ values
not too far from those which dominate the energy loss in finite length hot matter. The exact
region of validity of our approach should become apparent as we proceed. \par

Eq. (\ref{Diffusion1}) has a close resemblance to a Schr\"odinger, or diffusion, equation.
However, since the general Schr\"odinger equation cannot be solved explicitely this analogy
is not in itself useful. But, (\ref{Diffusion1}) is not only close to a Schr\"odinger
equation it is even close to the Schr\"odinger equation for a two dimensional harmonic
oscillator. To see this we recall that in infinite length matter the values of $B$ which
dominate the energy loss problem are those where $B^2 \sim \sqrt{\kappa}$ and we shall
later see the same to be true in finite length matter. 
Also, recall that 
$1 -\widetilde{V}(B^2) \approx {1 \over 4} B^2 \ln (1/B^2)$ 
for small $B$ for Coulomb scattering. 
At small $B^2$, $\ln (1/B^2)$ varies much more slowly than 
$B^2$ so we may expect a reasonable approximation to consist in neglecting 
the variation of $\ln (1/B^2)$ when solving (\ref{Diffusion1}). 
In this case (\ref{Diffusion1}) can be solved in terms of the general
solution to the Schr\"odinger equation for the 2-dimensional harmonic oscillator. 
In order to discuss the solution to (\ref{Diffusion1}) in a context somewhat 
more general than that of Coulomb scattering write this equation as
\beq
i {\partial \over \partial \tau} \widetilde{\vec{f_{\ }}}(\vec{B}, \tau ) = \left [
\widetilde{\kappa} \nabla_B^2 - i {B^2 \over 4} \widetilde{v}(B^2) \right ]
\widetilde{\vec{f_{\ }}} (\vec{B}, \tau ) \> ;
\qquad 
\widetilde{\vec{f}}(\vec{B},0)=
- \frac{i\pi}{2}\vec{B}\,\widetilde{v}(B^2)\>,
\label{Diffusion2} 
\eeq
where $\widetilde{v}(B^2) \equiv 4(1 - \widetilde{V}(B^2))/ B^2$. 
For potentials which decrease rapidly at large momentum transfer
$\widetilde{v}(B^2) \approx \widetilde{v}(0)$ at small $B^2$ and $\widetilde{v}(0)\mu^2$ is
just the mean momentum transfer squared. 
In the Coulomb case, $\widetilde{v}(B^2) \approx \ln 1/B^2$ for small
$B^2$ . We expect our approach to be valid whenever  \beq
B^2 {\partial \over \partial B^2} \ln \ \widetilde{v}(B^2) \ll 1 \ \ \ , \label{5.2}
\eeq  
for small $B^2$. In the Coulomb case we expect our answer for
$\omega \displaystyle{{dI / d \omega dz}}$ to be correct within logarithmic
accuracy in $1/\kappa$. 

So long as $\widetilde{v}(B^2)$ can be treated as a constant the solution to (\ref{Diffusion1})
proceeds in analogy with that of the two dimensional harmonic oscillator 
\beq
i {\partial \over \partial t} \psi (\vec{B}, t) = \left [ - {1 \over 2m}
\nabla_B^2 + {1 \over 2} m \omega_0^2 B^2 \right ] \psi (\vec{B}, t) \ \ \ .
\label{Oscillator} \eeq \noi Comparing (\ref{Diffusion2}) and (\ref{Oscillator}) it is clear
that one may take solutions to the harmonic oscillator if the identifications
\beq
m = - {1 \over 2 \widetilde{\kappa}} \quad , \quad \omega_0 = \sqrt{i
\widetilde{\kappa} \widetilde{v}} \label{Parameters} \eeq
\noi are made. Thus we may immediately write the function $\widetilde{\vec{f_{\ }}}$ in terms of
the Green function $G$ as 
\beq
\widetilde{\vec{f_{\ }}}(\vec{B}, \tau ) = \int d^2 \vec{B}' \ G(\vec{B}, \tau ;
\vec{B}', 0) \widetilde{\vec{f_{\ }}}(\vec{B}', 0) \ \ \ , \label{Green} \eeq
\noi where 
\beq
G(\vec{B}, \tau ; \vec{B}', 0) = {m \omega_0 \over 2 \pi i \sin \omega_0 \tau}
\exp \left \{ {i m \omega_0 \over 2 \sin \omega_0 \tau} \left [ (B^2 + B'^2) \cos
\omega_0 \tau - 2 \vec{B} \cdot \vec{B}' \right ] \right \} \ \ \ , \label{5.6} \eeq
\noi with $\omega_0$ and $m$ as given in (\ref{Parameters}). \par

Eq. (\ref{Green}) can be written as
\beq
\widetilde{\vec{f_{\ }}}(\vec{B}, \tau) = N \ e^{-\gamma B^2} \int d^2 \vec{B}' \
e^{-\alpha (\vec{B}' - \vec{C})^2} \ \widetilde{\vec{f_{\ }}} (\vec{B}', 0) \ \ \
, \label{fB}  \eeq \noi where
\beq
\alpha = {- i m \omega_0 \over 2 \tan \omega_0 \tau} \quad , \quad \gamma = {i
\over 2} m \omega_0 \tan \omega_0 \tau \quad , \quad \vec{C} = {\vec{B} \over
\cos  \omega_0 \tau} \quad , \quad N = {m \omega_0 \over 2 \pi i \sin \omega_0
\tau} \ .\label{Parameters2} \eeq
\noi Using $\sqrt{i} = \displaystyle{{1 + i \over \sqrt{2}}}$ one sees that ${\rm Re} \
\alpha > 0$ and $|\alpha | \gg 1$ for all values of $\tau$. Indeed for $\tau \ll
|\omega_0|^{-1}$ 

\bminiG{q-def5}
\alpha \approx {i \over 4 \widetilde{\kappa} \tau} + {1 \over 12} \widetilde{v} \tau \ \ \
,  \label{5.9a}  \eeeq
\noi while for $\tau \gg |\omega_0|^{-1}$
\beeq
\alpha \approx {1 + i \over 4} \sqrt{{\widetilde{v} \over 2 \widetilde{\kappa}}} \ \ \ .
\label{5.9b} 
\emini
Hence $|\alpha | \gsim 1/\sqrt{\widetilde{\kappa}}$ for all $\tau$. Thus so long as
$\widetilde{\kappa}$ is small the integral in (\ref{fB}) is dominated by values of $|\vec{B}'
-  \vec{C}|^2$ on the order of $|\alpha|^{-1} \ll 1$. Since we do not expect a rapid variation
of $\widetilde{\vec{f_{\ }}}(\vec{B}', 0)$ with $\vec{B}'$, the integral in (\ref{fB}) can be
done by replacing $\vec{B}'$ by $\vec{C}$ in that function and then carrying out the gaussian
integration over $\vec{B}'$. One finds
\beq
\widetilde{\vec{f_{\ }}}(\vec{B}, \tau ) = {\pi N \over \alpha} e^{-\gamma B^2} \ 
\widetilde{\vec{f_{\ }}}(\vec{C}, 0) \ \ \ . \label{5.10}
 \eeq
As we shall see below we always use values of $\tau$ and $\omega$ for which $\omega_0 \tau$
is of order 1. Thus from (5.8) it is clear that $\vec{C}$ is of the same order as
$\vec{B}$. \par

Recalling that
\beq
\widetilde{\vec{f_{\ }}}(\vec{C}, 0) = \widetilde{\vec{f_{\ }}}_0(\vec{C}) = - {\pi i
\over 2} \vec{C} \widetilde{v}(C^2) \ \ \ , \label{5.11} \eeq
we can write
\beq
\widetilde{\vec{f_{\ }}} (\vec{B}, \tau ) = - {i \pi \widetilde{v}(B^2) \vec{B} \over
2 \cos^2 \omega_0 \tau} \exp \left \{ - {i \over 2} m \omega_0 B^2 \tan \omega_0
\tau \right \} \ \ \ , \label{ftildevalue} 
\eeq 
where we have used the fact that $\widetilde{v}$ is slowly varying to replace $C^2$ by
$B^2$ as the argument of that function. Finally, from the above it is apparent that
$\vec{B}'$, in (\ref{fB}) is comparable to $\vec{B}$. This occurs because the parameter
$\widetilde{\kappa}$ in (\ref{Diffusion2}) is small thus severely limiting the diffusion
in $\vec{B}$ so that the time development of $\widetilde{\vec{f_{\ }}}(\vec{B}, \tau )$ occurs
with only moderate variations of $\vec{B}$. That is, the impact parameter of the produced
gluon is effectively frozen (at small values) during the whole time of its production. \par

We can now use (\ref{ftildevalue}) in (\ref{Bspec}). The integration over $\tau$ is made
simple by noting that 
\beq
\widetilde{\vec{f_{\ }}}(\vec{B}, \tau ) = {2 \pi i \vec{B} \over B^2} \ {\partial
\over \partial \tau} \exp \left \{ - {i \over 2} m \omega_0 B^2 \tan \omega_0
\tau \right \} \> . \label{5.13} 
\eeq
We emphasize that in the case when $\tilde{v}$ is a constant, (\ref{5.13})
is an \underline{exact} solution to (\ref{Diffusion2}).

Substituting (\ref{5.13}) into (\ref{Bspec}) leads to 
\beq
\left . \omega {dI \over d \omega dz} = -{3 \alpha_s C_R \over 4 \pi^2 \lambda_g \tau_0}
{\rm Re} \int_0^{\tau_0} d \tau \int \widetilde{v}(B^2) d^2\vec{B} \exp \left \{ - {i
\over 2} m \omega_0 B^2 \tan \omega_0 \tau \right \} \right
|_{\tilde{\kappa}}^{\tilde{\kappa} = 0} \ . \label{Bspec2} 
\eeq
We now do the $B$-integration, again neglecting the $B$-dependence of $\widetilde{v}$.
One finds 
\beq
\omega {dI \over d \omega dz} = {3 \alpha_s C_R \over 2 \pi \lambda_g \tau_0}
{\rm Re} \int_0^{\tau_0} d\tau 
\>2
\left [\, \frac{\omega_0}{\tan\omega_0 \tau} 
-  \frac{1}{\tau} \,\right]  . 
\label{Tauspec} 
\eeq
The $B$-integration in (\ref{Bspec2}) is determined by values of $B^2$ on the order of
$|m \omega_0 \tan \omega_0 \tau |^{-1}$. However, as is clear from (\ref{Tauspec}) values of
$|\omega_0\tau| \ll 1$ do not contribute significantly to 
$\omega \displaystyle{{dI / d\omega dz}}$,  
so we may assume that $|\omega_0 \tau | \geq 1$. 
This means that $B^2$ is on
the order of $|m \omega_0|^{-1}$ which, from (\ref{Parameters}) gives $B^2 \sim
\sqrt{\widetilde{\kappa}/\widetilde{v}(B^2)}$ as dominanting the spectrum 
of radiated gluons. In the Coulomb case this means 
$B^2 \sim \sqrt{\widetilde{\kappa}/\ln(1/\widetilde{\kappa})}$ 
while for potentials which decrease rapidly in transverse momentum, 
$B^2 \sim \sqrt{\widetilde{\kappa}}$  is dominant. 

The time integral in (\ref{Tauspec}) is easily performed and gives 
\beq
\omega{dI \over d \omega dz} = {6 \alpha_s C_R \over \pi L} \ \ln \left |
{\sin (\omega_0 \tau_0) \over \omega_0 \tau_0} \right | \> , \label{DiffSpectrum} 
\eeq 
where we have set $\tau_0 \widetilde{\lambda} = L$ the length of the matter.
The coupling $\alpha_s$ in (\ref{DiffSpectrum}) should be evaluated at momentum scale 
on the order of $\mu^2/B^2 = \mu^2 \sqrt{(\ln\ 1/\widetilde{\kappa})/\widetilde{\kappa}}$. 
We note that
$\omega \displaystyle{{dI/ d \omega dz}}$ is small when $|\omega_0 \tau_0| \ll 1$, 
while  
\beq
\omega{dI \over d \omega dz} \ 
\mathrel{\mathop {\longrightarrow}_{|\omega_0 \tau_0| \to \infty}} 
\ {3 \alpha_s C_R \over 2 \pi \lambda_g} \sqrt{2 \widetilde{\kappa}
\widetilde{v}(\sqrt{\widetilde{\kappa}})} \ \ \ . \label{5.17}  
\eeq
In the Coulomb case 
$\widetilde{v}(\sqrt{\widetilde{\kappa}}) = \half \ln (1/\widetilde{\kappa})$ 
bringing (\ref{5.17}) in agreement with (\ref{QCDBDMPS}) and with
(5.1) of BDMPS for infinite length matter.

\mysection{Energy loss in a finite length medium} \label{sec:nom6}
In order to find the total energy loss it is necessary to integrate (\ref{DiffSpectrum}) over
$\omega$. However, it is technically easier to go back to (\ref{Tauspec}) and do the
$\omega$-integral before doing the $\tau$-integration. Thus
\beq
- {dE \over dz} = \int \omega{dI \over d \omega dz} d\omega = {3 \alpha_s C_R
\over 2 \pi \lambda_g \tau_0} {\rm Re} \int_0^{\tau_0} {d \tau \over \tau}
\int_0^{\infty} d \omega 
\>2\left [\, {\omega_0 \tau \over  \tan (\omega_0 \tau)} - 1 \,\right ]  .
\label{lossa} 
\eeq
It is convenient to change variables taking
\beq
\widetilde{\kappa} = {\widetilde{\lambda} \mu^2 \over 2 \omega} \equiv {2 x^2
\over \tau^2 \widetilde{v}} \ \ \ . \label{xvariable} \eeq
\noi Then
\beq
d\omega = - {1 \over 2} \widetilde{\lambda} \mu^2 \tau^2 \widetilde{v} {dx \over
x^3} \ \ \ . \label{6.3} \eeq
Using (\ref{xvariable}) in (\ref{lossa})
\beq
- {dE \over dz} = {3 \alpha_s C_R \mu^2 \over \pi \tau_0} {\rm Re}
\int_0^{\tau_0} \widetilde{v} \tau d \tau \int_0^{\infty} {dx \over x^3} \left \{
{(1 + i)x \over \tan [(1 + i)x]} - 1 \right \} \ \ \ . \label{lossb} 
\eeq
Using (see Appendix D)
\beq
{\rm Re} \int_0^{\infty} {dx \over x^3} \left [ {(1 + i)x \over \tan [(1 + i)x]} -
1 \right ] = {\pi \over 6} \ \ \ , \label{PiIntegral} \eeq
\noi we finally arrive at
\beq
- {dE \over dz} = {\alpha_s C_R \over 4} \mu^2 \tau_0 \ \widetilde{v} (1 / \tau_0)
= {\alpha_s C_R \over 4} \ {\mu^2 \over \widetilde{\lambda}} L \ \widetilde{v}(1/
\tau_0) \ \ \ . \label{Loss1} 
\eeq
In determining the argument of $\widetilde{v}$ we note that $x$ is of order 1 in
(\ref{lossb}) while $\tau$ is of order $\tau_0$. From (\ref{xvariable}) we have that
$\widetilde{\kappa}$ is of order $[\tau_0^2 \widetilde{v}]^{-1}$ in the dominant region of
integration giving $B^2$ of order $[\tau_0 \widetilde{v}]^{-1}$ as the effective
argument of $\widetilde{v}$. 
Thus in the Coulomb case and with logarithmic accuracy
\beq
- {dE \over dz} = {\alpha_s C_R \over 8} \ {\mu^2 \over \lambda_g} L \ 
\ln\frac{L}{\lambda_g} \>. \label{Loss2} 
\eeq
The total energy loss in the matter is just $L$ times (\ref{Loss2}) giving
\beq
 -\Delta E = {\alpha_s C_R \over 8} \ {\mu^2 \over \lambda_g}  L^2 \ 
\ln \frac{L}{\lambda_g} \> . \label{TotalLoss} 
\eeq
We note here, as was noted earlier by BDPS, that the energy loss depends on the hot
matter only through the parameter $\mu^2/\lambda$ in the logarithmic approximation. The
fact that the total energy loss scales as $L^2$ is remarkable and, to our knowledge, has not
been anticipated by earlier discussions in the literature. This loss in energy is due to
emission of gluons having typical energy 
\beq
\omega = (\widetilde{\lambda} \tau )^2 \ {\mu^2 \over \widetilde{\lambda}} \ 
{\widetilde{v} \over 4x^2} \approx {\mu^2 \over \lambda_g} \ {L^2 \ 
\widetilde{v}(\lambda_g/L) \over 160} \ \ \ . \label{TypicalEnergy} 
\eeq
In arriving at (\ref{TypicalEnergy}) we have taken $\tau = \tau_0/ \sqrt{2}$ 
as the median value of $\tau$ in (\ref{lossb}) and we have set $x \approx 3$ 
(see \ref{D.5}) as the median value of $x$ in (\ref{PiIntegral}). 

Of course any estimate of energy loss in possibly realistic circumstances in heavy ion
collisions is hazardous and should be received with caution and scepticism. Nevertheless,
we feel it necessary to say something about the size of the results (\ref{Loss2}) and
(\ref{TotalLoss}). For hot QCD matter having temperature $T = 250$ MeV, an
estimate \cite{Peigne}, using simple perturbative formulas, gives $\lambda \approx 1\fm$ and
$\mu^2/\widetilde{\lambda} \approx 1 \GeV/\fm^2$. 
For $L= 10 \fm$  we find from
(\ref{TotalLoss}) that $\Delta E = 80 \alpha_s\> \GeV$ 
while for 
$L = 5 \fm$,  $\Delta E = 20\alpha_s\> \GeV$, which are rather large numbers if 
$\alpha_s \gsim 1/3$--$1/2$ or so. Note also
that, from (\ref{TypicalEnergy}), typical values of $\omega$ for $L = 10 \fm$ are of order 
3 \GeV, which means that several soft gluons are emitted in a typical event.

\vspace{1.5 cm}
\noindent
{\bf\large Acknowledgement}

This research is supported in part by the EEC Programme ``Human Capital
and Mobility'', Network ``Physics at High Energy Colliders'',
Contract CHRX-CT93-0357.

\newpage
\appendix
\mysection{Radiation amplitude induced by two scatterings}
\label{radiation}
As a means of producing an energetic quark we consider deep inelastic scattering where the
quark is produced at a time $t_0$. If $t_0 = - \infty$ we have the circumstance considered
by BDPS and BDMPS, while $t_0 = 0$ corresponds to a quark produced in the medium. 

The radiation amplitude is given by the sum of the seven diagrams shown below, with $t_1$
and $t_2$ the interaction times of the incident quark or of the radiated gluon with static
centres 1 and 2. The emission time of the gluon is denoted by $t$. The radiation amplitude
is simply calculated in the framework of time-ordered perturbation theory. For each
diagram, we indicate the associated phase factor and the interval over which the emission
time $t$ has to be integrated. The phase is given by the energy difference between 
the states
after and before the interaction. We then exhibit the resulting difference of phase
factors, the corresponding kinematical factors involving transverse momenta (see
(\ref{M1bis}) and (\ref{M2bis})) obtained as in the case of a single scattering,
together with the associated colour structure. A colour factor $(T^b T^a T^c)_{B'B}$ will be
denoted as $bac$. We also factor out the elastic double scattering cross section.
\bea
\begin{minipage}[cr]{.33\textwidth}
\includegraphics[width=\textwidth]{fig14a.ps}
\vspace{10pt}
\end{minipage} 
& \propto & \int_{t_0}^{t_1} dt \ e^{it {k_{\bot}^2 \over 2 \omega}} \nn \\
\Rightarrow   \left (
e^{it_1 {k_{\bot}^2 \over 2 \omega}} - e^{it_0 {k_{\bot}^2 \over 2 \omega}}  \right ) 
 {\vec{k}_{\bot} \over k_{\bot}^2}  \ bac & \ &  \label{A.1} \eea
\bea
\begin{minipage}[cr]{.33\textwidth}\includegraphics[width=\textwidth]{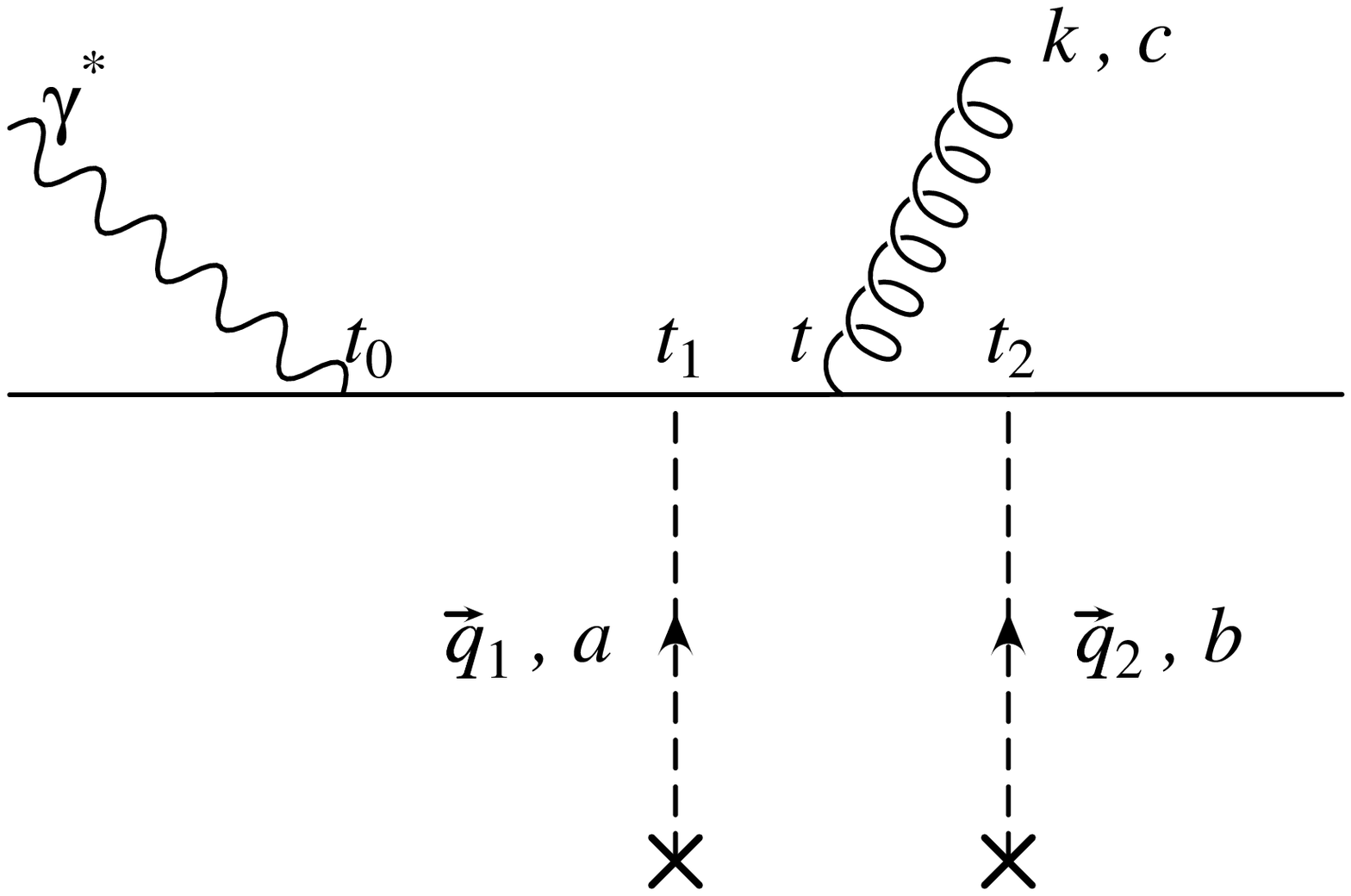}
\vspace{10pt}
\end{minipage} 
& \propto & \int_{t_1}^{t_2} dt \ e^{it {k_{\bot}^2 \over 2 \omega}} \nn \\
\Rightarrow   \left (e^{it_2 {k_{\bot}^2 \over 2 \omega}} - e^{it_1 {k_{\bot}^2 \over 2 \omega}}
 \right ) {\vec{k}_{\bot} \over k_{\bot}^2} \ bca  & \ &  \label{A.2} \eea
\bea
\begin{minipage}[cr]{.33\textwidth}
\includegraphics[width=\textwidth]{fig14c.ps}
\vspace{10pt}
\end{minipage} 
& \propto & \int_{t_2}^{\infty} dt \ e^{it {k_{\bot}^2 \over 2 \omega}} \nn \\
\Rightarrow  - e^{it_2 {k_{\bot}^2 \over 2 \omega}} {\vec{k}_{\bot} \over k_{\bot}^2} \ cba   & \ &  \label{A.3} \eea
\bea
\begin{minipage}[cr]{.33\textwidth}
\includegraphics[width=\textwidth]{fig14d.ps}
\vspace{10pt}
\end{minipage} 
\propto  \int_{t_1}^{t_2} dt \ e^{it {(k - q_2)_{\bot}^2 \over 2 \omega}+it_2
{k_{\bot}^2 - (k - q_2)_{\bot}^2 \over 2 \omega}} & \  & 
\begin{minipage}[c]{.2\textwidth}
\end{minipage} \nn \\
\Rightarrow  \left ( e^{it_2 {k_{\bot}^2
\over 2 \omega}} - e^{it_1 {(k-q_2)_{\bot}^2 \over 2 \omega} + it_2 {k_{\bot}^2 - (k -
q_2)_{\bot}^2 \over 2 \omega}} \right ) {\vec{k}_{\bot} - \vec{q}_{2 \bot} \over (\vec{k} -
\vec{q}_2)_{\bot}^2} \ [c,b]a & \ &  \label{A.4} \eea
\bea
\begin{minipage}[cr]{.33\textwidth}
\includegraphics[width=\textwidth]{fig14e.ps}
\vspace{10pt}
\end{minipage} 
\propto  \int_{t_0}^{t_1} dt \ e^{it {(k - q_1 - q_2)_{\bot}^2 \over 2 \omega}+it_1
{(k - q_2)_{\bot}^2 - (k - q_1 - q_2)_{\bot}^2 \over 2 \omega} + it_2 {k_{\bot}^2 - (k -
q_2)_{\bot}^2 \over 2 \omega}} \nn \\
\Rightarrow  e^{it_2 {k_{\bot}^2 - (k - q_2)_{\bot}^2 \over 2 \omega}} \left ( e^{it_1
{(k-q_2)_{\bot}^2 \over 2 \omega}} - e^{it_1 {(k-q_2)_{\bot}^2 -(k - q_1 -
q_2)_{\bot}^2 \over 2 \omega} + it_0 {(k - q_1 - q_2)_{\bot}^2 \over 2 \omega}}
\right ) {\vec{k}_{\bot} - \vec{q}_{1\bot} - \vec{q}_{2\bot} \over (\vec{k}_{\bot} -
\vec{q}_1 - \vec{q}_2)_{\bot}^2} \ [[c,b],a] \label{A.5} \eea
\bea
\begin{minipage}[cr]{.33\textwidth}
\includegraphics[width=\textwidth]{fig14f.ps}
\vspace{10pt}
\end{minipage} 
\propto \int_{t_0}^{t_1} dt \ e^{it {(k - q_1)_{\bot}^2 \over 2 \omega}+it_1
{k_{\bot}^2 - (k - q_1)_{\bot}^2 \over 2 \omega}}  \nn \\
\Rightarrow \left ( e^{it_1 {k_{\bot}^2
\over 2 \omega}} - e^{it_1 {k_{\bot}^2 - (k - q_1)_{\bot}^2 \over 2 \omega} + it_0
{(k - q_1)_{\bot}^2 \over 2 \omega}} \right ) {\vec{k}_{\bot} - \vec{q}_{1 \bot} \over
(\vec{k} - \vec{q}_1)_{\bot}^2} \ b[c,a] \label{A.6} \eea
\newpage 
\bea
\begin{minipage}[cr]{.33\textwidth}
\includegraphics[width=\textwidth]{fig14g.ps}
\vspace{10pt}
\end{minipage} 
\propto \int_{t_0}^{t_1} dt \ e^{it {(k - q_2)_{\bot}^2 \over 2 \omega}+it_2
{k_{\bot}^2 - (k - q_2)_{\bot}^2 \over 2 \omega}}  \nn \\
\Rightarrow  e^{it_2 {k_{\bot}^2
- (k - q_2)_{\bot}^2 \over 2 \omega}} \left ( e^{it_1 {(k-q_2)_{\bot}^2 \over 2 \omega}} -
e^{it_0 {(k-q_2)_{\bot}^2 \over 2 \omega}} \right ) {(\vec{k}_{\bot} - \vec{q}_{2})_{\bot}
\over (\vec{k} - \vec{q}_2)_{\bot}^2} \ a[c,b] \label{A.7} \eea

When we take the production time to be $t_0 = - \infty$, exponential factors containing
$t_0$ disappear and the radiation amplitude induced by two scatterings takes the simple form
given in (\ref{Mrad1}).

 \mysection{Diagrammatic calculation of colour factors}
\label{diagrammatic}
The generators $T^a$ ($a = 1, \ldots N_c^2 - 1$) of the fundamental representation of
$SU(N_c)$ satisfy
\bminiG{q-def6}
[T^a, T^b] &=& i \ f^{abc} \ T^c \ \ \ , \label{B.1a} 
\eeeq
and
\beeq
\left \{ T^a , T^b \right \} &=& {1 \over N_c} \delta^{ab} + d^{abc} \ T^c \ \ \ ,  
\label{B.1b} 
\emini
where the structure constant $f^{abc}$ and the symbol $d^{abc}$ 
are respectively totally antisymmetric and symmetric.

\noi Using
\bminiG{q-def7}
Tr (T^a T^b) &=& T_R \ \delta^{ab} = {1 \over 2} \delta^{ab} \label{B.2a} \ \ \ , \\
(T^a T^a )_{ik} &=& C_F \ \delta_{ik} = {N_c^2 - 1 \over 2N_c} \delta_{ik}
\label{B.2b} \ \ \ , \\
f^{acd} f^{bcd} &=& C_A \ \delta^{ab} = N_c \ \delta^{ab} \ \ \ ,
\label{B.2c} 
\emini
\noi together with the decomposition
\beq
T^a T^b = {1 \over 2} [T^a, T^b] + {1 \over 2} \left \{ T^a, T^b \right \} \ \ \
, \label{B.3} \eeq
\noi and the relations \citd{Muta}{Macfarlane}
\bminiG{q-def8}
T^b T^a T^b = - {1 \over 2N_c} T^a \ \ \ , \label{B4.a} \\
f^{aib} \ f^{bjc} \ f^{cka} = - {N_c \over 2} f^{ijk} \ \ \ , \label{B4.b}
\emini
\noi we get the following simple diagrammatic rules:
\beq
\centering
\begin{minipage}[c]{.25\textwidth}
\includegraphics[width=\textwidth]{fig15a.ps}
\end{minipage}
\begin{minipage}[c]{.15\textwidth}
$\ \ \ \ =\ \ {1 \over 2}\ \ $ 
\end{minipage}
\begin{minipage}[c]{.25\textwidth}
\includegraphics[width=.8\textwidth]{gluon.ps}
\end{minipage}
\eeq
\vskip 1cm
\beq
\centering
\begin{minipage}[b]{.25\textwidth}
\includegraphics[width=\textwidth]{fig15b.ps}
\end{minipage}
\begin{minipage}[c]{.15\textwidth}
$\ \ \ \ =\ \ C_F\ \ $ 
\end{minipage}
\begin{minipage}[c]{.25\textwidth}
\includegraphics[width=.8\textwidth]{quark.ps}
\end{minipage}
\eeq
\vskip 1cm
\beq
\centering
\begin{minipage}[c]{.25\textwidth}
\includegraphics[width=\textwidth]{fig15c.ps}
\end{minipage}
\begin{minipage}[c]{.15\textwidth}
$\ \ \ \ =\ \ N_c\ \ $
\end{minipage}
\begin{minipage}[c]{.25\textwidth}
\includegraphics[width=.8\textwidth]{gluon.ps}
\end{minipage}
\eeq
\vskip 1cm
\beq
\centering
\begin{minipage}[b]{.25\textwidth}
\includegraphics[width=\textwidth]{fig15d.ps}
\end{minipage}
\begin{minipage}[c]{.15\textwidth}
$\ \ \ \ =\ \ {N_c \over 2}\ \ $
\end{minipage}
\begin{minipage}[b]{.25\textwidth}
\includegraphics[width=.8\textwidth]{qqgvertex.ps}
\end{minipage}
\eeq
\vskip 1cm
\beq
\centering
\begin{minipage}[c]{.25\textwidth}
\includegraphics[width=\textwidth]{fig15e.ps}
\end{minipage}
\begin{minipage}[c]{.15\textwidth}
$\ \ =\ \ {-\frac{1}{2 N_c}}$
\end{minipage}
\begin{minipage}[b]{.25\textwidth}
\includegraphics[width=.8\textwidth]{qqgvertex.ps}
\end{minipage}
\eeq
\vskip 1cm
\beq
\centering
\begin{minipage}[c]{.25\textwidth}
\includegraphics[width=\textwidth]{fig15f.ps}
\end{minipage}
\begin{minipage}[c]{.15\textwidth}
$\ \ \ \ =\ \ {N_c  \over 2}\ \ $
\end{minipage}
\begin{minipage}[b]{.25\textwidth}
\includegraphics[width=.8\textwidth]{gggvertex.ps}
\end{minipage}
\eeq
\vskip 1cm

\mysection{Expression of the medium induced spectrum}
\label{expression}
We start from the first term of (\ref{Spectrum2}) and calculate the product of currents given
there and in (\ref{fig10}). After averaging over the trajectory of the incident quark, the
contribution of the interference $\vec{J}_i \cdot \vec{J}_j \ e^{i(\varphi_{_i} -
\varphi_{_j})}$ will depend only on $(j - i)$. Thus we simply evaluate this product for $i =
1$ and $j = n+2$, where $n = j - i - 1$ is the number of intermediate gluon rescatterings.

Since the probability density $V(q^2)$ is isotropic, we can choose the momentum convention
separately for graphs $Y$ and $H$ as shown in (\ref{fig10}). 
We separate the colour and momentum structure of these graphs as
\beq
Y + H = Y_C Y_M + H_C H_M \ \ \ .  \label{C.1}
\eeq
\noi We get 
\beq
Y_M = \mu^{-2} \vec{J}_1 \cdot \vec{J}_{n+2} = - H_M \ \ \ , \label{C.2}
\eeq
\noi where the dimensionless current $\vec{J}_i$ is given in (\ref{kinCurrent}) in terms of
the dimensionless momenta (\ref{Rescaling}), and, from Appendix B, 
\beq
Y_C = {N_c (N_c/2)^{n+1} N_c \ C_F \over N_c \ C_F^{n+2}} = N_c \left ( {N_c \over 2 C_F}
\right )^{n+1} = - 2 H_C \ \ \ , \label{C.3}
\eeq
\noi where the factor $N_c C_F^{n+2}$ in the denominator accounts for the normalization to
the multiple elastic scattering cross section. \par

Finally the phases associated with $Y$ and $H$ are equal,
\bea
\varphi (Y) &=& \mu^2 \left ( t_1 {U_1^2 \over 2 \omega} + \sum_{\ell = 2}^{n+2} t_{\ell}
{U_{\ell}^2 - U_{\ell - 1}^2 \over 2 \omega} \right ) - \mu^2 \left ( t_{n+2} {U_{n+2}^2
\over 2 \omega} \right ) \nn \\
&=& - \kappa \sum_{\ell = 1}^{n+1} \ {t_{\ell + 1} - t_{\ell} \over \lambda} \ U_{\ell}^2 =
\varphi (H) \ \ \ . \label{C.4}
\eea 

Collecting (\ref{C.1}) to (\ref{C.4}) we get the large-$N_c$ formula (\ref{LPM}).

\mysection{A curious integral}
\label{integral}
Here we evaluate the integral (\ref{PiIntegral}),
\bminiG{D.1}
I &=& \int_0^{\infty} g(x) \ dx \ \ \ , \label{D.1.a} \\
g(x) &=& {\rm Re} {1 \over x^3} \left [ {(1 + i)x \over \tan [(1 + i)x]} - 1 \right ] 
= {1\over x^3} \left [ x {{\rm sh} \ 2x + \sin 2x \over {\rm ch} \ 2x - \cos 2x} 
- 1 \right ] \ \ \ ,  \label{D.1.b} 
\emini
\noi 
by using the integration contour in the complex $z = x + iy$ plane as shown in Fig.~11.

\begin{figure}[h]
\label{fig16}
\centering
\includegraphics[width=.5\textwidth]{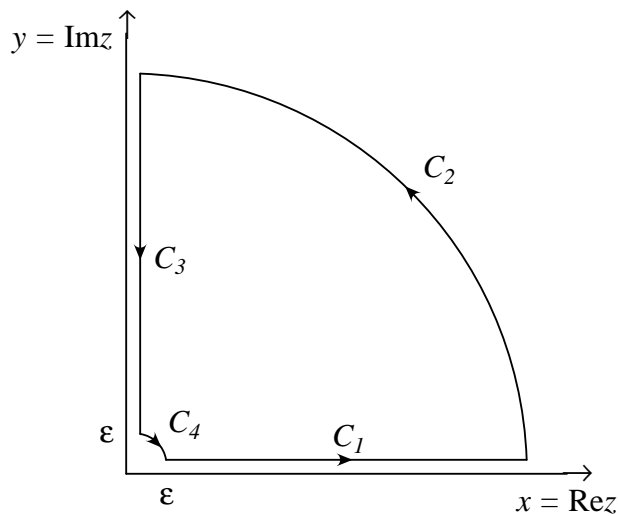}
\caption{\it{Integration contour for the integral (\ref{D.1}).}}
\end{figure}

\noi Adding the contributions from the paths $C_1$ and $C_3$ gives for $\varepsilon \to 0$
\beq
I_{C_1} + I_{C_3} = 2 \lim_{\varepsilon \to 0} \int_{\varepsilon}^{\infty} g(x) \
dx = 2I \ \ \ . \label{D.2}
\eeq
\noi When closing the path at infinity the contribution along $C_2$ vanishes, whereas due
to the single pole of the integrand at $z = 0$ we have
\beq
I_{C_4} = - {2 \pi i \over 4} \left ( - {1 \over 3} \right ) (1 + i)^2 = - {\pi \over 3}
\ \ \ . \label{D.3} 
\eeq
Observing that the poles of $1/\tan [(1\! +\! i)z]$, 
$z_n = \half(1 - i) \pi n$ are not in this quadrant we find the result given 
in (\ref{PiIntegral}) for $I$, 
\beq
I = {1 \over 2} \left ( I_{C_1} + I_{C_3} \right ) = - {1 \over 2} I_{C_4} = {\pi \over 6}
\label{D.4} \>. 
\eeq
Finally, we evaluate numerically the median value of $x$ in (\ref{D.1.a}) defined by
\bminiG{D.5}
\int_0^{x_{med}} g(x) \ dx \>=\>\frac12 \, I \>=\> \frac{\pi}{12}\>.
\eeeq
We find 
\beeq
x_{med} \>\simeq \> 3.2\> .
\emini

\newpage
\vbox to 2 truecm {}
\def\labelenumi{[\arabic{enumi}]}
\noindent
{\bf\large References}
\ben
\item\label{LP}
  L.D.~Landau and I.Ya.~Pomeranchuk,
  {\em Dokl.~Akad.~Nauk SSSR}\/  \underline {92} (1953) 535, 735.

\item\label{Mig}
  A.B.~Migdal, \pr{103}{1811}{56} ; and references therein.

\item\label{Ter}
  M.L.~Ter-Mikaelian, {\it High Energy Electromagnetic Processes
in Condensed Media}, John Wiley \& Sons, NY, 1972.

\item\label{SLAC}
P.L.~Anthony et al., \prl{75}{1949}{95}; \\ S.~Klein et al., preprint SLAC-PUB-6378 T/E (November 1993).

\item\label{BD} R.~Blankenbecler and S.D.~Drell, \prD{53}{6265}{96}.

\item\label{GW} M.~Gyulassy and X.-N.~Wang, \np{420}{583}{94}.

\item\label{GW2} M.~Gyulassy, X.-N.~Wang and M.~Pl\"{u}mer, \prD{51}{3236}{95}.

\item\label{BDPS} R.~Baier, Yu.L.~Dokshitzer, S.~Peign\'e and D.~Schiff,
\pl{345}{277}{95}.

\item\label{BDMPS}
 R.~Baier, Yu.L.~Dokshitzer, A.H.~Mueller, S.~Peign\'e and D.~Schiff,
LPTHE-Orsay 95-84 (February 1996).

\item\label{BDMPS3} R.~Baier, Yu.L.~Dokshitzer, A.H. Mueller, S.~Peign\'e and D.~Schiff,
(in preparation).

 \item\label{AM} A. H. Mueller, in Proceedings of Workshop on Deep Inelastic Scattering and
QCD, Paris, April 1995, Eds. J.-F. Laporte and Y. Sirois, p.~29.  

\item\label{Muta} T. Muta, {\it Foundations of Quantum Chromodynamics}, World Scientific
Lecture Notes in Physics, Vol.~5. 

\item\label{Macfarlane} A. J. Macfarlane, A. Sudbery et P. H. Weisz, {\it Commun. Math.
Phys.} \underbar{11} (1968) 77.

\item\label{Peigne} S. Peign\'e, Ph. D. thesis, Paris-Sud University, May 1995.
\een
 \end{document}